\documentclass[a4paper,fleqn,usenatbib]{mnras}

\usepackage{amsmath} 
\usepackage{amssymb} 
\usepackage{multirow}

\usepackage{graphicx}
\usepackage{grffile}
\usepackage[dvips]{epsfig}
\usepackage{epsfig}  
\usepackage{color}
\usepackage{caption}
\usepackage{hyperref}
\usepackage{bm}

\def\be{\begin{equation}}
\def\ee{\end{equation}}
\def\ba{\begin{eqnarray}}
\def\ea{\end{eqnarray}}

\definecolor{red}{rgb}{1,0.0,0.0}

\definecolor{darkgreen}{rgb}{0.0,0.5,0.0}

\newcommand{\lya}{\ifmmode{{\rm Ly}\alpha}\else Ly$\alpha$\ \fi}
\newcommand{\cm}{\ifmmode{{\rm cm}}\else cm\fi}

\newcommand{\ergps}{\,{\rm erg}\,{\rm s}\ifmmode{}^{-1}\else ${}^{-1}$\fi}
\newcommand{\Mpch}{\,{\rm Mpc}\,\ifmmode h^{-1}\else $h^{-1}$\fi}

\newcommand{\kms}{\ifmmode\mathrm{km\ s}^{-1}\else km s$^{-1}$\fi}
\newcommand{\hMsun}{{\ifmmode{h^{-1}{\rm{M_{\odot}}}}\else{$h^{-1}{\rm{M_{\odot}}}$}\fi}}
\newcommand{\Msun}{{\ifmmode{{\rm{M_{\odot}}}}\else{${\rm{M_{\odot}}}$}\fi}}

\begin{document}

\title[LG satellites distribution asphericity]{We are not the 99 percent: quantifying
  asphericity in the distribution of Local Group satellites}
\author[J.E. Forero-Romero \& V. Arias]
{Jaime E. Forero-Romero $^{1}$ \thanks{je.forero@uniandes.edu.co},
Ver\'onica Arias$^1$\\
$^1$ Departamento de F\'isica, Universidad de los Andes, Cra. 1
  No. 18A-10 Edificio Ip, CP 111711, Bogot\'a, Colombia \\
}

\maketitle

\begin{abstract}
We use simulations to build an analytic probability distribution for
the asphericity in the satellite distribution around Local Group (LG)
type galaxies in the Lambda Cold Dark Matter (LCDM) paradigm. 
We use this distribution to estimate the  atypicality
of the satellite distributions in the LG even when the underlying
simulations do not have enough systems fully resembling the LG in
terms of its typical masses, separation and kinematics.
We demonstrate the method using three different simulations
(Illustris-1,  Illustris-1-Dark and ELVIS) and a number of satellites
ranging from 11 to 15.
Detailed results differ greatly among the simulations suggesting a
strong influence of the typical DM halo mass, the number of satellites
and the simulated baryonic effects.   
However, there are three common trends.
First, at most $2\%$ of the pairs are expected to have satellite
distributions with the same asphericity as the LG; second,
at most $80\%$ of the pairs have a halo with a satellite
distribution as aspherical as in M31; and third, at most $4\%$ of the
pairs have a halo with satellite distribution as planar as in the MW. 
These quantitative results place the LG at the level of a $3\sigma$
outlier in the LCDM paradigm. 
We suggest that understanding the reasons for this atypicality
requires quantifying the asphericity probability distribution as a
function of halo mass and large scale environment.
The approach presented here can facilitate that kind of study and other
comparisons between different numerical setups and choices to study
satellites around LG pairs in simulations.  
\end{abstract}


\section{Introduction}

The spatial distribution of satellite galaxies around our Milky Way
(MW) and the M31 galaxy is becoming a stringent test for structure
formation theories in an explicit cosmological context. 
This started with the suggested existence of a Magellanic Plane, a flattened
structure of satellite galaxies and globular clusters around the MW,
by \cite{1976RGOB..182..241K} and \cite{1976MNRAS.174..695L}.  
Forty years later \cite{2005A&A...431..517K} quantified that the
highly planar distribution of the 11 classical MW satellites has less
than a $0.5\%$ chance to happen by chance if the
parent distribution is spherically symmetric, interpreting this as a
challenge to the Lambda Cold Dark Matter (LCDM) paradigm.
The same year it was recognized, using numerical simulations, that
this comparison was unfair given that Dark Matter halos in LCDM
are expected to be triaxial and not spherical.
Nevertheless, some estimates of the chances to find simulated satellites as
planar as in the MW 
were as low as $2\%$ \citep{2005ApJ...629..219Z,2005A&A...437..383K} while others expected
planar satellite configurations in every DM halo
\citep{2005MNRAS.363..146L}.

Later, \cite{2007MNRAS.374.1125M} used simulations to confirm the low
chances ($<0.5\%$) found in \cite{2005A&A...431..517K}.
This result was challenged by \cite{2009MNRAS.399..550L} who continued
to report high chances to find a planar configuration;
two more recent numerical experiments with high resolution simulations
(\cite{2013MNRAS.429..725S} and \cite{2016MNRAS.457.1931S}) reported
contradicting results (low and high chances to find the observational
result, respectively) but precise figures were not quoted in these three reports. 
Meanwhile, \cite{2013MNRAS.429.1502W} and \cite{2014ApJ...789L..24P} published
the results of new tests using high resolution simulations and cosmological
volumes arriving at a chance of $6\%$ and $0.77\%$, respectively, to
have a satellite distribution as triaxial as the observations.

In the case of M31, all studies agree on the point that the spatial
distribution of the $15$ to $27$ brightest satellites  are consistent with
spherically symmetric distributions and easy to reproduce in LCDM simulations
\citep{2006AJ....131.1405K,2007MNRAS.374.1125M, 2013ApJ...766..120C}.
In a very different observational statement, there is a subset of $15$
satellites out of the total population of $27$ satellites that show a
highly planar distribution
\citep{2013ApJ...766..120C,2013Natur.493...62I,2015ApJ...799L..13C}.
This so-called satellite plane is deemed to be difficult to produce in
LCDM simulations and is thought to be transient structure
\citep{2015MNRAS.452.3838C,2015ApJ...809...49B,2015ApJ...800...34G,2016MNRAS.460.4348B,2017MNRAS.466.3119A,2017MNRAS.465..641F}.  
However, in this paper we do not address these \emph{plane finding}
algorithms and focus instead on the characterization of the global
satellite distribution ranked either by decreasing luminosity, maximum
circular velocity or stellar mass. 
Table \ref{table:M31} and Table \ref{table:MW} summarize some more
details on the results we have just mentioned \footnote{
Those tables also include the results from this paper using the
methodology we describe in the upcoming sections.}.

\begin{table*}
\centering
\begin{tabular}{|p{4.0cm}|p{4.5cm}| p{5.5cm}| c|}\hline
Reference & Target Measurement & Parent Simulation & Probability ($\%$)\\\hline
\text{\cite{2006AJ....131.1405K}} & RMS width in 15 brightest
satellites & Monte Carlo satellite distributions with a power law
radial distribution & $87-99$\\
\text{\cite{2007MNRAS.374.1125M}} & $c/a$ ratio and RMS width in 16 brightest satellites & Monte Carlo from a spherical power law radial
distribution & $17$\\
\text{\cite{2013ApJ...766..120C}}& RMS width in the 27 brightest
satellites & Monte Carlo randomized satellite distribution & High\\
This Work & $c/a$ ratio, $b/a$ ratio and RMS width in 11-15 brightest
satellites & 27 halos from selected pairs in a cosmological N-body Dark Matter only simulation ($\sim
10^{6}$\Msun particle mass resolution)& $<80$ \\
This Work & $c/a$ ratio, $b/a$ ratio and RMS width in 11-15 brightest
satellites & 24 halos from selected pairs in a cosmological N-body hydro simulation ($\sim
10^{6}$\Msun particle mass resolution)& $<71$ \\
This Work & $c/a$ ratio, $b/a$ ratio and RMS width in 11-15 brightest
satellites & 12 high resolution DM only N-body simulations ($\sim
10^{5}$\Msun particle mass resolution) of halo pairs & $<57$ \\
\hline
\end{tabular}
\caption{Probability to find  the triaxiality and/or root mean squared
  (RMS) width of M31 satellites in LCDM simulations. 
\label{table:M31}}
\end{table*}

\begin{table*}
\centering
\begin{tabular}{|p{4.0cm}|p{4.5cm}| p{5.5cm}| c|}\hline
Reference & Target Measurement & Parent Simulation & Probability ($\%$)\\\hline
\text{\cite{2005A&A...431..517K}} & RMS width in 11 classical
satellites & Monte Carlo from a spherical power law
radial distribution. & $<0.5$ \\
\text{\cite{2005MNRAS.363..146L}} & $c/a$ ratio of 11 classical
satellites & 6 high resolution DM only N-body simulations ($\sim10^5$
\Msun\ particle mass resolution). & High\\ 
\text{\cite{2005ApJ...629..219Z}} & $c/a$ ratio and disk width in 
11 classical satellites & 3 high resolution DM only N-body
simulations. ($\sim10^6$ \Msun\ particle mass resolution) & 2 \\
\text{\cite{2007MNRAS.374.1125M}} & $c/a$ ratio and RMS width in 11-13 brightest satellites & Monte Carlo from a halo triaxiality distribution from LCDM
simulations & $<0.5$\\
\text{\cite{2009MNRAS.399..550L}}& $c/a$ ratio in 11 classical satellites & 436 halos from a
cosmological N-body simulation ($1.3\times 10^{8}$\Msun\ particle mass)
&  High \\
\text{\cite{2013MNRAS.429..725S}}& $c/a$ ratio in 12 brightest
satellites & 6 High resolution DM only N-body simulations ($\sim
10^3$\Msun particle mass resolution) of individual halos & Low \\
\text{\cite{2013MNRAS.429.1502W}}& $c/a$ ratio or RMS width in 11 brightest
satellites & 1686 halos from a cosmological DM only N-body simulation
($\sim 10^6$\Msun particle mass resolution) & 6 or 13 \\
\text{\cite{2014ApJ...789L..24P}}& $c/a$ and $b/a$ ratio in 11
classical satellites & 48 high resolution DM only N-body simulations
($\sim 10^{5}$\Msun particle mass resolution) of both halo pairs and
isolated halos & 0.77\\
\text{\cite{2016MNRAS.457.1931S}}& $c/a$ ratio in 11 classical satellites & 12
high resolution Hydro simulation of halo pairs ($\sim 10^{4}$\Msun particle
mass resolution) & High\\
This Work & $c/a$ ratio, $b/a$ ratio and RMS width in 11-15 brightest
satellites & 27 halos from selected pairs in a cosmological N-body
Dark Matter only simulation ($\sim 10^{6}$\Msun particle mass
resolution)& $<4$\\
This Work & $c/a$ ratio, $b/a$ ratio and RMS width in 11-15 brightest
satellites & 24 halos from selected pairs in a cosmological N-body
hydro simulation ($\sim 10^{6}$\Msun particle mass resolution)& $<1.5$ \\
This Work & $c/a$ ratio, $b/a$ ratio and RMS width in 11-15 brightest
satellites & 12 high resolution DM only N-body simulations ($\sim
10^{5}$\Msun particle mass resolution) of halo pairs & $<1.6$\\
\hline
\end{tabular}
\caption{Same as Table \ref{table:M31} for the MW satellites.
\label{table:MW}}
\end{table*}

Some of the difficulty in trying to reconcile and understand the
seemingly conflicting or inconclusive results on the MW has its origin on
the frequentist fashion generally used to compute probabilities.
Usually, this process starts by building a high level parent sample in the
simulation and then counting how many elements has the subset
meeting some criteria. 
This has two inconveniences.
The first is that the probability estimate is made against whatever
turns out to be typical in each simulation. 
A fair comparison across simulations would require first
characterizing all the simulations at the high level parent samples,
something that is difficult to do in practice. 
The second inconvenient is that the systems that fully resemble
the LG (i.e. its stellar mass content, morphology or kinematics) have
a low cosmological number density \citep{ForeroRomero2013}, this means
that for the current cosmological volumes in simulations the high
level parent sample has a small size, making it hard to derive robust
probabilities by counting.    
In this paper we present and demonstrate a method to overcome these two
limitations.

We control the first effect by setting as a direct point of reference
a spherical satellite distribution and not the simulations themselves.
The spherical satellite distribution is built from the data itself
(observational or simulated) by randomizing the angular position
of each satellite around the central galaxy and keeping its radial
distance fixed \citep{2017AN....338..854P}. 
We characterize the satellites in terms of the scalars describing its
deviation from the spherical distribution. 
We then build an analytic probability distribution for the
asphericity; this solves the problem of having a common reference
point to compare simulations.  
We use a simulation to estimate the parameters in this distribution to
later use it as a parent sample to generate any desired number of
samples that are by construction statistically compatible with the
simulation; thus overcoming the problem of having a small number of
systems in the parent simulations. 

To summarize, we use asphericity to characterize on equal footing
simulations and observations.  
Then, we build an analytical probability distribution  for the
asphericity and use simulations to estimate its free parameters.  
Finally, we use these distributions to generate large numbers of
samples and directly estimate the number of systems meeting a desired
set of criteria.

The rest of the paper describes in detail our implementation and
results. It is structured as follows. 
In Section \ref{sec:DataSamples} we list the sources of the observational and
simulated data to be used throughout the paper.
In Section \ref{sec:SpatialMeasurements} we describe the methods we
use to build halo pairs and  
characterize their satellite distributions.
In Section \ref{sec:results} we present our results to finally
conclude in Section \ref{sec:conclusions}.

\section{Data samples}\label{sec:DataSamples}

\subsection{Observational Data}
\label{sec:obs}

The base for our analysis is the catalog compiled by
\cite{2014yCat..74351928P} which reports information on all 
galaxies within 3 Mpc around the Sun to that date. 
Detailed description of the compiled catalog can be found in
\cite{2013MNRAS.435.1928P}, here we summarize the relevant features
for the current study.
The information in the catalogue is based on the catalogue compiled by
\cite{2012AJ....144....4M}.
The distance estimates are based on resolved stellar populations. 
We use three dimensional positions in a cartesian coordinate system
as computed by \cite{2013MNRAS.435.1928P}.
In this coordinate system the $z$-axis points towards the Galactic north pole, the
$x$-axis points in the direction from the Sun to the Galactic center,
and the $y$-axis points in the direction of the Galactic rotation.

For both the M31 and MW we only use the 11 to 15 brightest satellites (using
$M_V$ magnitudes) within a distance of $300$kpc from its central galaxy.
The satellites included for the MW analysis are: 
LMC, SMC, Canis Major, Sagittarius dSph, Fornax, Leo I, Sculptor,
Leo II, Sextans I, Carina, Ursa Minor, Draco, Canes Venatici (I),
Hercules and Bootes II.
The satellites included for the M31 analysis are: Triangulum, NGC205,
M32, IC10, NGC185, NGC147, Andromeda VII, Andromeda II, Andromeda
XXXII, Andromeda XXXI, Andromeda I, Andromeda VI, Andromeda XXIII, LGS 3, 
 and Andromeda III.

\subsection{Simulated Data}
 
We use data from three different simulations: Illustris-1,
Illustris-1-Dark and ELVIS. 
In what follows we describe the relevant detailed of those
simulations, how the halo pairs resembling the LG are selected and how
the satellite samples for each halo is built.
Figure \ref{fig:physical_pairs} summarizes the physical properties (maximum
circular velocities, radial velocities and separation) of the halo
pairs to be used.

\subsubsection{Illustris-1 and Illustris-1-Dark}
\label{sec:illustris}

We use publicly available data from the Illustris Project 
\citep{2014MNRAS.444.1518V}. 
This suite of cosmological simulations were performed using the quasi-Lagrangian
code AREPO \citep{2010MNRAS.401..791S}.
They followed the coupled evolution of dark  matter and gas and
includes parametrizations to account for the effects of 
gas cooling, photoionization, star formation, stellar feedback, black
hole and super massive black hole feedback. 
The simulation volume is a cubic box with a $75$ \Mpch\ side.
The cosmological parameters correspond to a $\Lambda$CDM cosmology
consistent with WMAP-9 measurements \citep{2013ApJS..208...19H}. 

We extract halo and galaxy information from the Illustris-1 and
Illustris-1-Dark simulations. 
The former includes hydrodynamics and star formation prescriptions while the latter only
includes dark matter physics. 
These simulations have the highest resolution in the current release of the
Illustris Project.
Illustris-1 has $1820^3$ dark matter particles and $1820^3$ initial gas
volume elements, while Illustris-1-Dark has $1820^3$ dark matter particles.
This corresponds to a dark matter particle mass of
$6.3\times 10^6$\Msun\ and a minimum mass for the baryonic volume
element of $1.2\times 10^6$\Msun\ for Illustris-1 and a dark matter
particle mass of $7.6\times 10^6$\Msun\ for Illustris-1-Dark.
In both simulations the dark matter gravitational softening is $1.4$
kpc.

We build a sample of pairs that resemble the conditions in the LG as follows.
First, we select from Illustris-1 all the galaxies with a stellar mass
in the range $1\times10^{10}\Msun <M_{\star}<1.5 \times 10^{11} \Msun$.
Then we select the pairs with the following conditions.

\begin{itemize}
\item For each galaxy $A$ we find its closest galaxy $B$, if galaxy $A$ is also
the closest to $B$, the two are considered as a pair. 
\item With $d_{AB}$ the distance between the two galaxies and
  $M_{\star,min}$ the lowest stellar mass in the two galaxies, we
  discard pairs that have any other galaxy $C$ with stellar mass
  $M_{\star}>M_{\star, min}$ closer than $3\times d_{AB}$ from any of
  the pair's members. 
\item The distance $d_{AB}$ is greater than $700$ kpc.
\item The relative radial velocity between the two galaxies, including
  the Hubble flow, is $-120\ \kms <v_{AB,r}<0\ \kms$. 
\end{itemize}

We find 27 pairs with these conditions. 
In Illustris-1-Dark we use the center of mass position of the 27 pairs
in Illustris-1 to find the matching halo pairs.
After discarding the pairs with less than 11 detected subhalos in one
of the halos we end up with a total of 24 pairs in Illustris-1-Dark. 
This corresponds to a pair number density of $\sim 2 \times10^{-5}$
pairs Mpc$^{-3}$. 
Figure \ref{fig:physical_pairs} summarizes the physical properties (maximum
circular velocities, radial velocities and separation) of the halo
pairs to be used.

\subsection{Data from the ELVIS project}
\label{sim:ELVIS}

We use data from the public release of the Exploring the Local
Universe In Simulations (ELVIS) project.
For a detailed description of that project and its data we refer the
reader to \cite{2014MNRAS.438.2578G}. 
Here we summarize the elements relevant to our discussion.

ELVIS data comes from resimulations of dark matter halo pairs selected
in dark matter only cosmological simulations. 
The parent cosmological boxes have a cosmology consistent with the
Wilkinson Microwave Anisotropy Probe 7 results.
The ELVIS project used the results from $50$ simulation boxes of side
length $70.4$ Mpc to select pairs with kinematic characteristics
similar to the LG. 
These selection criteria included the following
\begin{itemize}
\item The virial mass of each host must be in the range 
$1\times
  10^{12} M_{\odot}< M_{vir}<3\times 10^{12}M_{\odot}$ 
\item The total pair mass must be in the range
$2\times
  10^{12} M_{\odot}< M_{vir}<5\times 10^{12}M_{\odot}$ 
\item The center of mass separation is in the range $0.6\leq d\leq1$
  Mpc.
\item The relative radial velocity is negative.
\item No halos more massive than the least massive halo within $2.8$
  Mpc and no halos with $M_{vir}>7\times 10^{13}$ within $7$ Mpc of
  the pairs' center of mass.
\end{itemize} 

This corresponds to a number density of $\sim 8 \times10^{-6}$
pairs/Mpc$^{3}$, which is a factor $\sim 2.5$ lower than the pair
number density we find in the Illustris-1 data.
There were a total of 146 pairs that met those criteria, but only $12$
were chosen for resimulation. 
Additionally, the selected pairs for resimulation have a relative
tangential velocity less than $75 $\kms. 
The dark matter particle resolution in these resimulations is
$1.9\times 10^5$, an order of magnitude times better than Illustris-1.
In this paper we only use the results from these $12$ resimulated pairs.

\section{Building, Characterizing and Comparing Satellites Spatial Distributions}
\label{sec:SpatialMeasurements}

\subsection{Building Satellite Samples}

We compare the joint satellite distributions in the MW and M31 at fixed
satellite number, $N_s$.
We make this choice to explicitly model the influence of satellite numbers
on the statistics. 

Although Illustris-1 has stellar particles, we do not use their
properties to select the satellite population for the results reported
here. 
The smallest galaxies are barely resolved in stellar mass at
magnitudes of $M_V=-9$, close to the limit of the 11 ``classical'' MW
satellites, we use instead the dark matter information. 

We decide instead to use the same satellite selection rules in
Illustris-1, Illustris-1-Dark and ELVIS.
Namely, selecting the satellites by ranking the subhalos in decreasing
order of its current maximum circular velocity and select the first
$N_s$ halos in the list. 

These selection criteria also  allow us to perform a fair comparison
across simulations. 
In the case of Illustris-1 versus Illustris-1-Dark, it let us direct
our attention at baryonic physics as the driver behind the differences
in the results between these two simulations.  
To a good approximation these satellites are the systems with the
highest circular velocity at infall \citep{2011MNRAS.415L..40B}, which
is the physical quantity expected to best correlate with luminosity
\citep{2004ApJ...609...35K,2006ApJ...647..201C,2010MNRAS.404.1111G}.  
With this choice the maximum circular velocity cut corresponding to the
smallest satellite included in the sample is different for each case;
in terms of resolution the subhalos are resolved with at least
$\approx 40 (400)$ particles for the Illustris (ELVIS) simulation.

We compute the satellite statistics for $11\leq N_s\leq 15$.
The lower limit corresponds to the number of classical MW satellites,
while the upper limit corresponds to the minimum number of
satellites usually included in M31 studies.
This boundary in the satellite number is also present due to the
limits in the resolution of Illustris.

In observations we rank the satellites by its $M_V$
magnitude.  
We also tested using $M_V$ and stellar mass in Illustris-1, and the
stellar mass in ELVIS (provided through abundance matching) as a
selection criterion. 
Our main conclusions remain unchanged by those choices. 

\begin{figure*}
\centering
\includegraphics[width=0.30\textwidth]{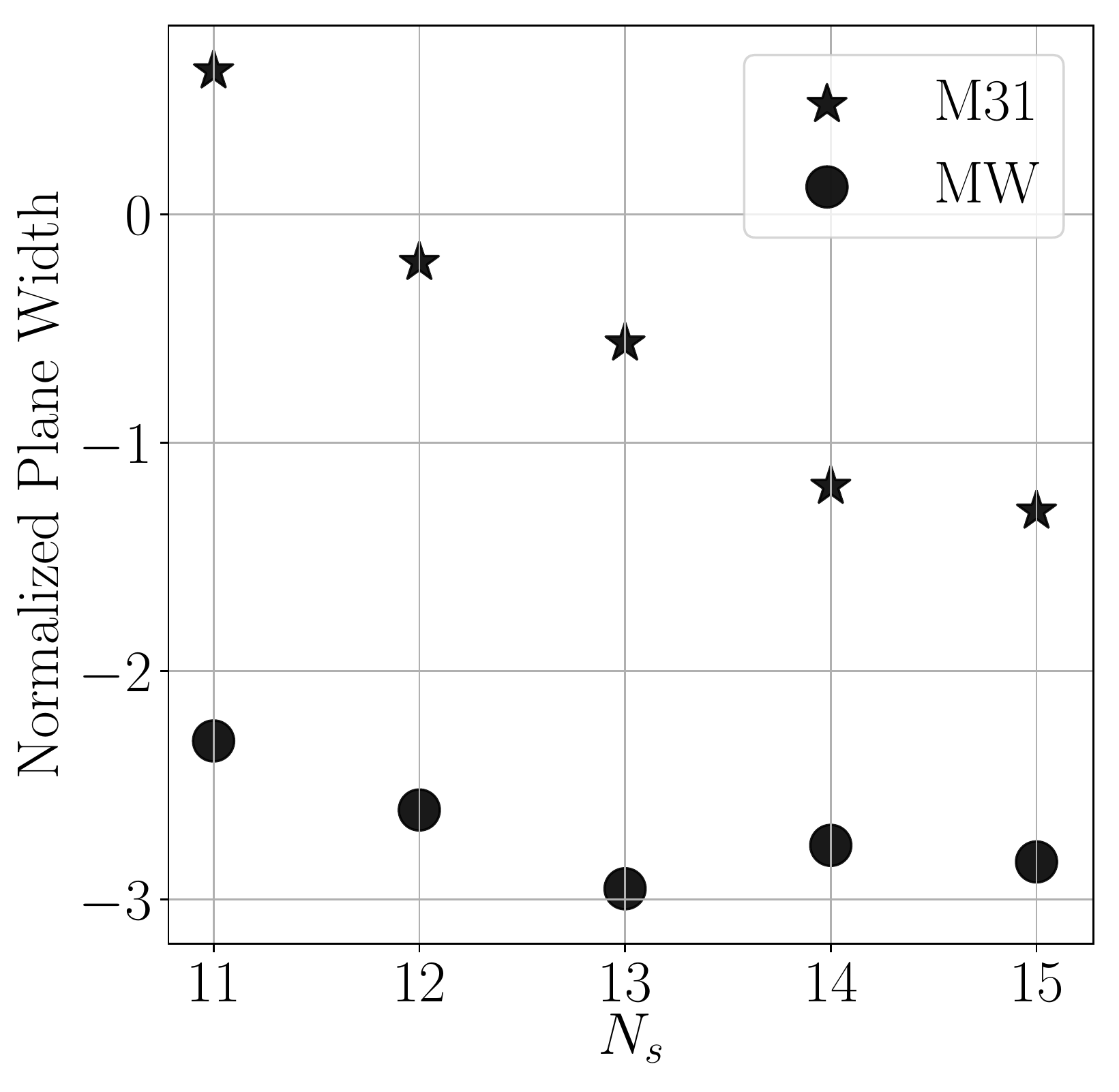}
\includegraphics[width=0.30\textwidth]{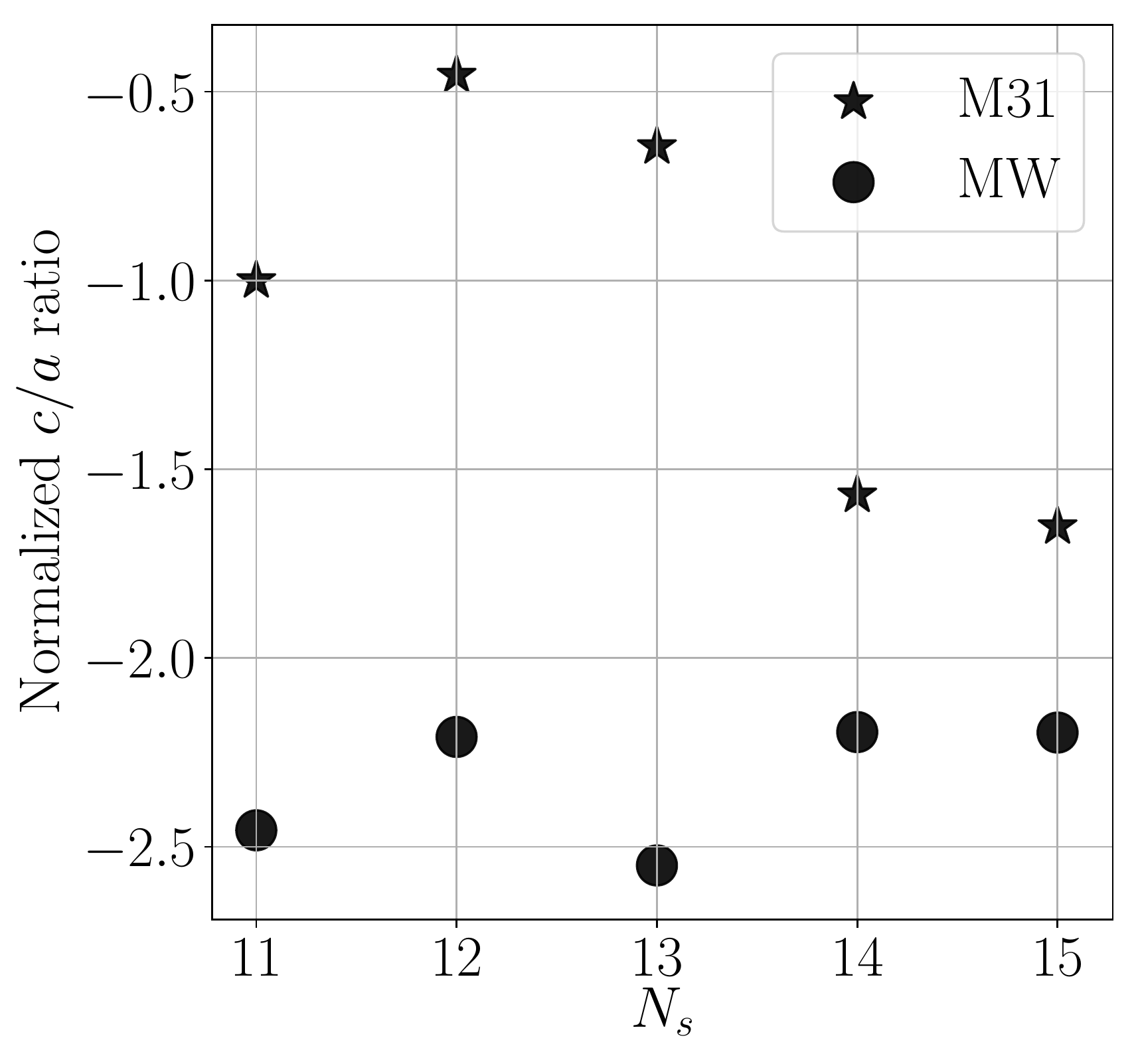}
\includegraphics[width=0.30\textwidth]{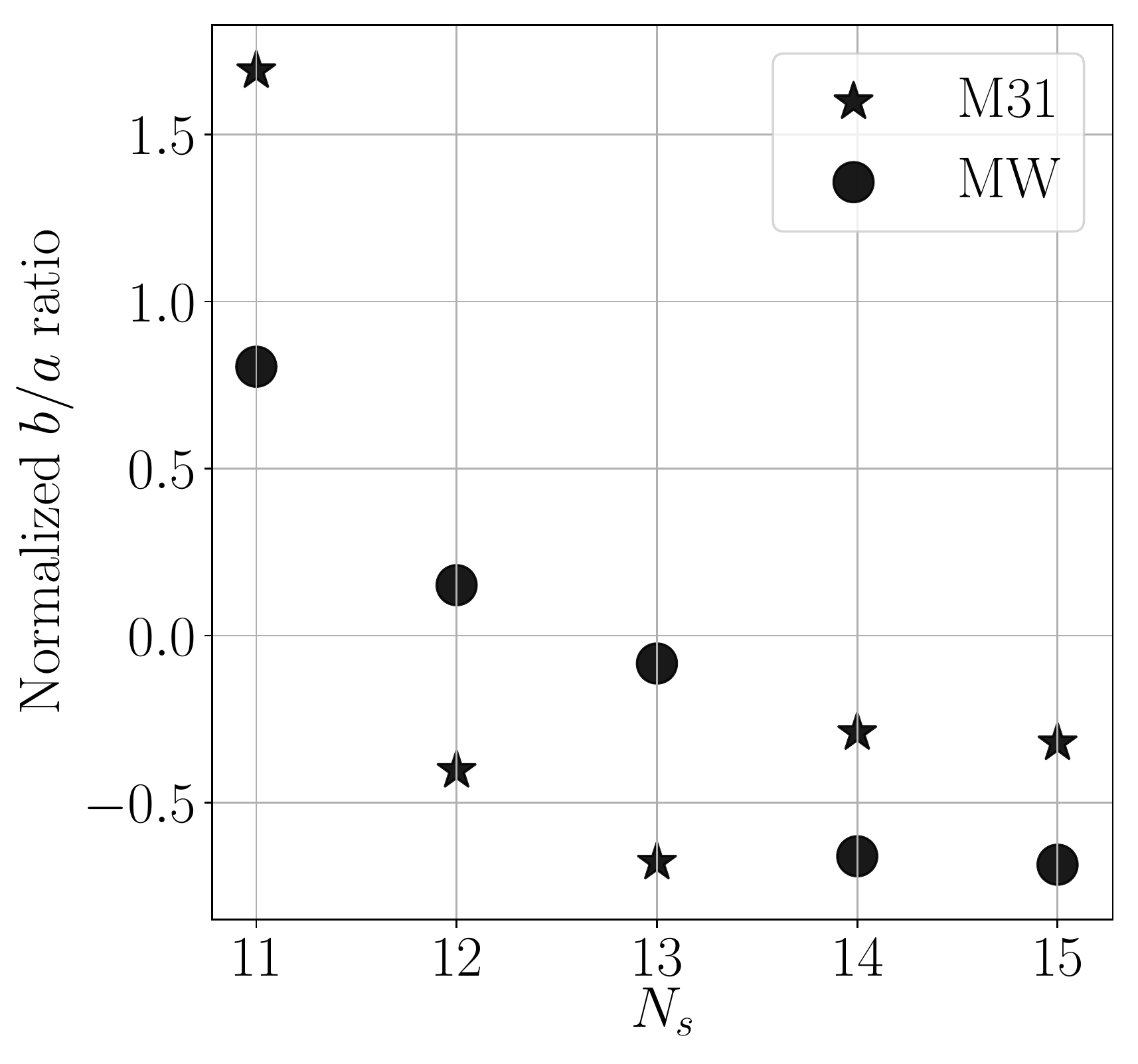}
\caption{Asphericity scalars from observations as a function
  of satellite number. 
  Left: plane width; center: $c/a$ ratio;
  right: $b/a$ ratio. 
  These quantities can take positive or negative values because they
  were recentered to the mean and normalized to the standard deviation 
  of the spherically randomized satellites as explained in Section
  \ref{sec:normalization}. 
  The plane width and $c/a$ ratio for the MW are
  consistently found \emph{beyond} two standard
  deviations of the expected result for an statistically spherical
  distribution. 
  M31 shows the opposite trend and are always found
  \emph{within} two standard deviations. \label{fig:normalized_n}} 
\end{figure*}

\subsection{Describing Samples with the Inertia Tensor}

We describe the satellites with the inertia
tensor defined by the satellites' positions.  

\begin{equation}
{\bf{\bar{I}}} = \sum_{i=1}^{N_p}[(\bf{r}_i - \bf{r}_0)^2\cdot \bf{1} -
  (\bf{r}_i-\bf{r}_0)\cdot (\bf{r}_i - \bf{r}_0)^{T}],
\label{eq:intensor}
\end{equation}
where $k$ indexes the set of satellites of interest
$\bf{r}_k$ are the satellites' positions, $\bf{r}_{0}$ is the location
of the central galaxy $\bf{1}$ is the unit matrix, and  
${\bf r}^T$ is the transposed vector $\bf{r}$. 
We use $\bf{r}_0$ as the position of the central galaxy, and not the
satellites' geometrical center, to allow for a fair comparison once
the angular positions of the satellites are randomized around this
point. 

From this tensor we compute its eigenvalues,
$\lambda_1>\lambda_2>\lambda_3$, and corresponding eigenvectors,
$\hat{I}_1$, $\hat{I}_2$, $\hat{I}_3$.
We define the size of the three ellipsoidal axis as
$a=\lambda_1$, $b=\lambda_2$ and $c=\lambda_3$.
We also define $\hat{n}\equiv \hat{I}_1$ as the vector perpendicular to the
planar satellite distribution. 
We also define the Root Mean Squared (RMS) plane width $w$ as the
standard deviation of the satellite distances to the plane defined by
the vector $\hat{n}$.    

To summarize we characterize the satellite distribution with the following
quantities obtained from the inertia tensor: 
\begin{itemize}
\item RMS plane width, $w$.
\item $c/a$ axis ratio.
\item $b/a$ axis ratio.
\end{itemize}

\subsection{Characterizing asphericity with normalized scalars}
\label{sec:normalization}
We compare each satellite distribution against its own spherically
randomized distribution.
We keep fixed the radial position of every satellite
with respect to the central galaxy and then randomize its angular
position. 
We repeat this procedure $10^4$ times for each satellite distribution
and proceed to measure the quantities mentioned in the previous section:
$w$, $c/a$ and $b/a$.
This allows us to build a normalized version of all quantities of
interest by subtracting the mean and dividing by the standard
deviation of the values from randomized samples.
We use the normalized quantities to build the analytic probability
distributions for the scalars describing asphericity.

\subsection{Building an analytic asphericity probability distribution}

After building the normalized variables for the simulated data we perform a
Kolmogorov-Smirnov test with the null hypothesis of belonging to a
normal distribution with mean and standard deviation computed from the
mean and standard deviation estimated from the data. 
Although the physical quantities of interest are bound, we find that the
distributions for the normalized $w$, $c/a$ and $b/a$ are indeed
consistent with gaussian distributions. 

Based on this result we build a multivariate normal distribution for
the joint distributions of the normalized $w$, $c/a$ and $b/a$:

\begin{equation}
p(X; \mu, \Sigma) = \frac{1}{(2\pi)^{3/2}|\Sigma|^{1/2}}
\exp\left(-\frac{1}{2}(X-\mu)^{T}\Sigma^{-1}(X-\mu)\right), 
\label{eq:multivariate}
\end{equation}
where $X=[w, c/a, b/a]^{T}$ is a vector variable with the normalized
quantities, $\mu$ is the vector mean and the $\Sigma$ is the
covariance matrix.

We compute the preferred covariance matrix and the mean distribution values
with a jackknife technique. 
That is, out of the $n$ pairs in each simulation, we perform $n$
different covariance and mean value measurements using only $n-1$ pairs. 
The reported covariance and mean values correspond to the average of
all measurements, the corresponding standard deviation also helps us to
estimate the uncertainty on every reported coefficient.
This compact description allows us to generate samples of size $N$
that are consistent by construction with their parent simulation. 

Finally, we use the generated samples to estimate how
common are the deviations from sphericity that we measure in the
observational data.  
We use a double-tailed test around the mean in this comparison,
meaning that we count the fraction of generated points points with
absolute values larger than the absolute observed value with the mean
substracted.


\begin{figure*}
\centering
\includegraphics[width=0.48\textwidth]{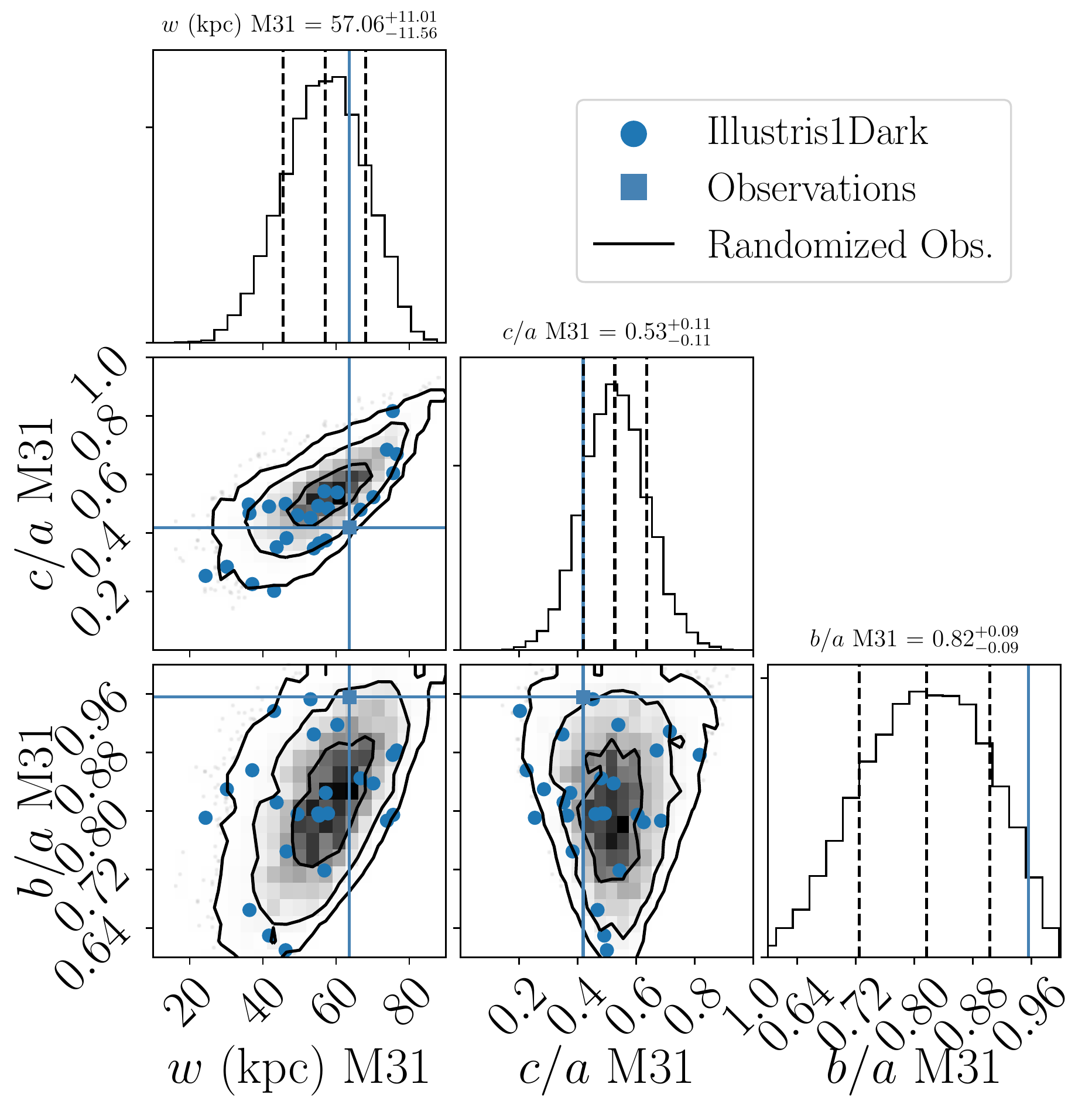}
\includegraphics[width=0.48\textwidth]{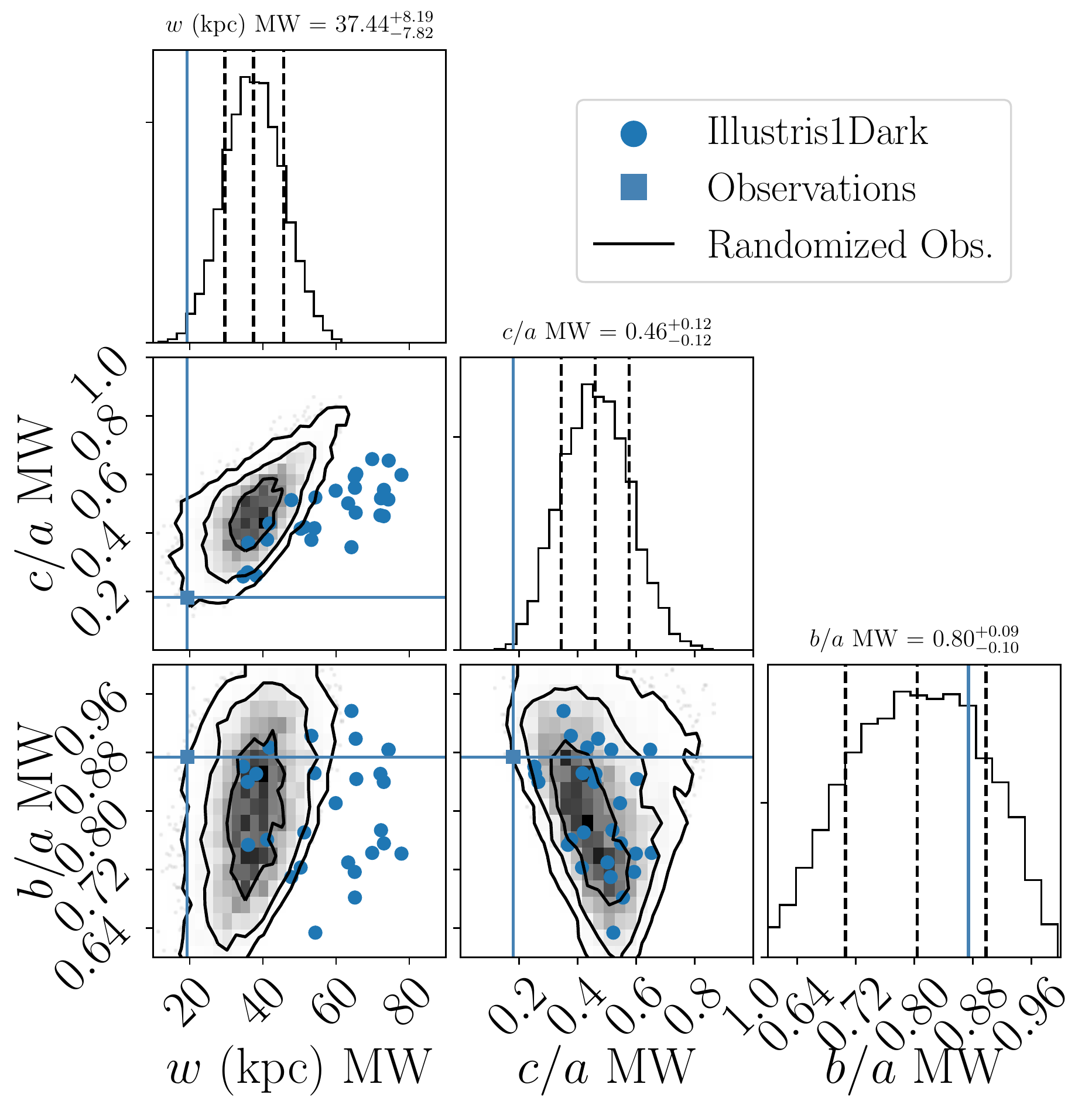}
\caption{Scalars in physical units describing satellite shape around the LG galaxies for
  a fixed number of satellites $N_s$=11: 
  plane width $w$, $c/a$ ratio and $b/a$ ratio.
  Left: M31, right: MW. 
  The histograms along the diagonal show the distribution for the
  $10^4$ values after spherical randomization; the vertical dashed
  lines show the $16^{\rm th}$, $50^{\rm th}$ and $84^{\rm th}$
  percentiles of this distribution; the vertical solid line correspond
  to observations. 
  Outside the diagonal there are 2D histograms showing the ($1$, $2$,
  $3$) $\sigma$ contours for the randomized data.  
  The square at the center of the cross are observations.
  The blue circles are the results from Illustris-1-Dark. 
  For the M31 these three results
  (observations, randomized observations and simulations) seem to be
  in broad agreement, while for the MW they clearly differ. 
  The corresponding plots for Illustris-1 and ELVIS are
  in Figures \ref{fig:all_plots_illustris1} and
\ref{fig:all_plots_elvis}, respectively. \label{fig:physical_illustris1dark}}
\end{figure*}

\begin{figure*}
\centering
\includegraphics[width=0.48\textwidth]{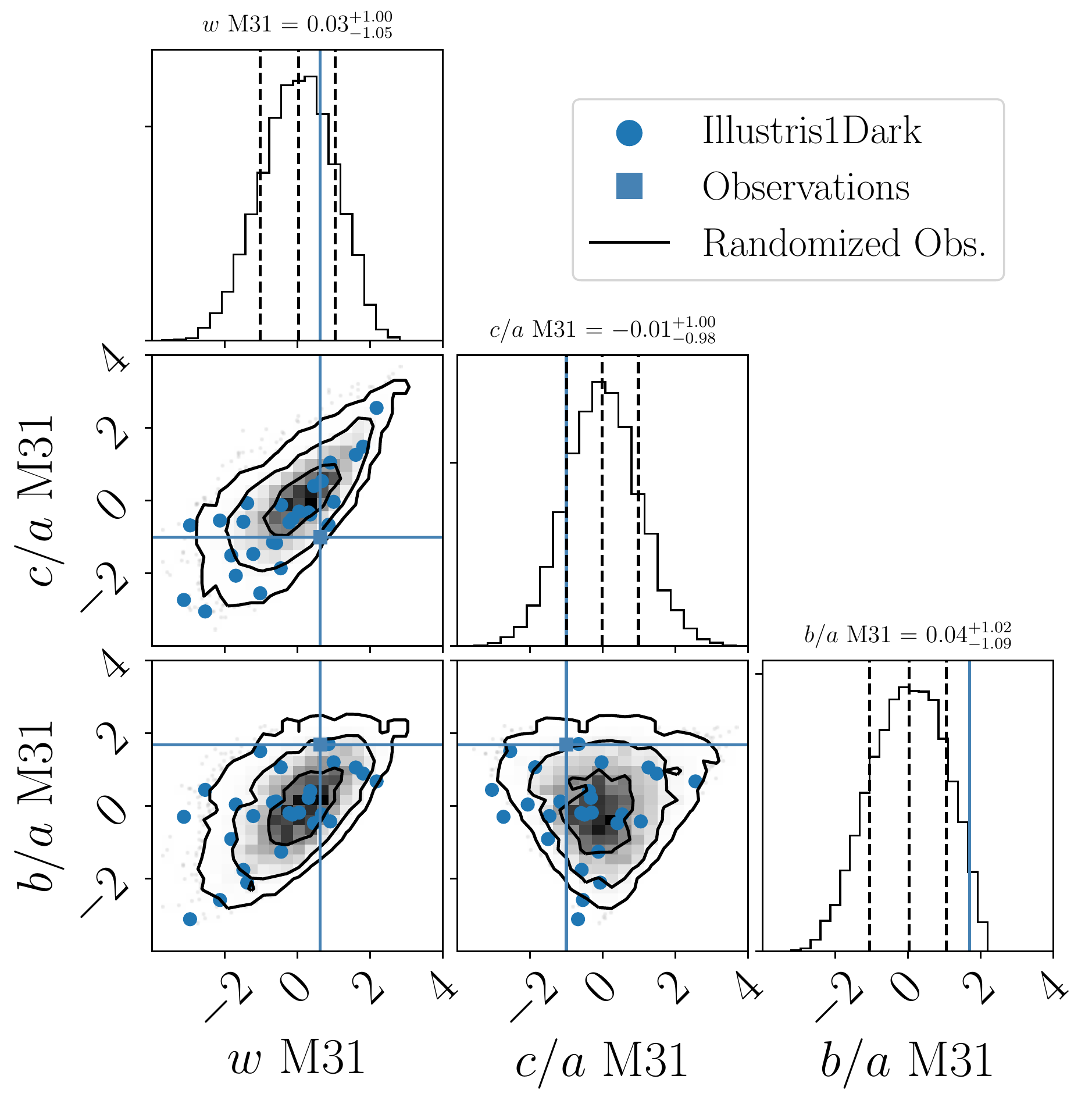}
\includegraphics[width=0.48\textwidth]{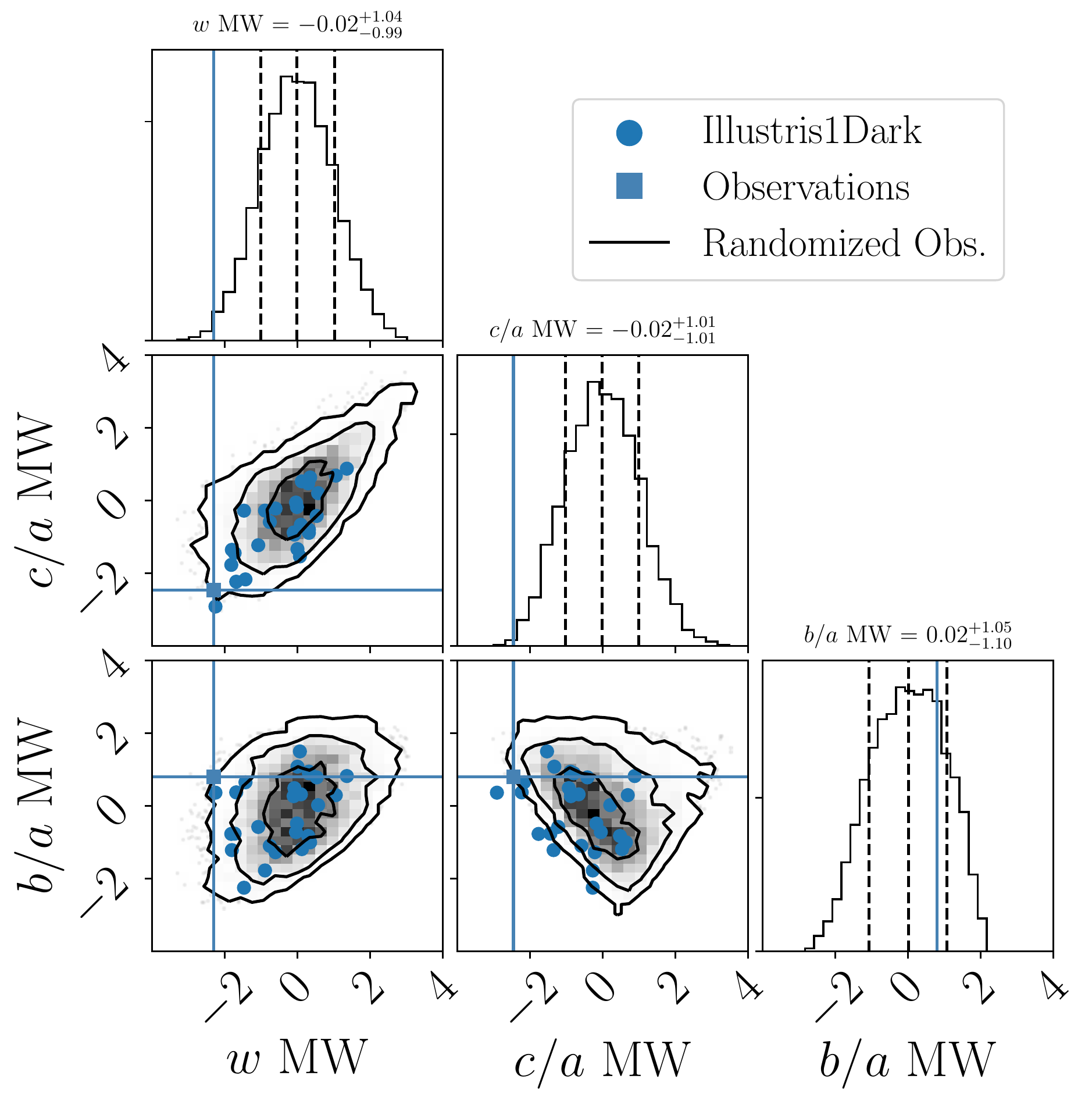}
\caption{
Same layout as Figure \ref{fig:physical_illustris1dark}. 
  Each scalar
  quantity is re-centered and normalized by the mean value and standard
  deviation from the $10^4$ randomizations. 
  The distribution from the randomizations (plots over the diagonal)
  now has by definition a mean of zero and standard deviation of one. 
The results for the width and $c/a$ ratio of MW satellites are more
than two standard deviations away from the mean of the randomized
data, both in the 1D distributions (diagonal plots) and the 2D
distributions (off-diagonal plots). 
The corresponding plots for Illustris-1 and ELVIS are
in Figures \ref{fig:all_plots_illustris1} and
\ref{fig:all_plots_elvis},
respectively. \label{fig:normalized_illustris1dark}}     
\end{figure*}

\begin{figure*}
\centering
\includegraphics[width=0.45\textwidth]{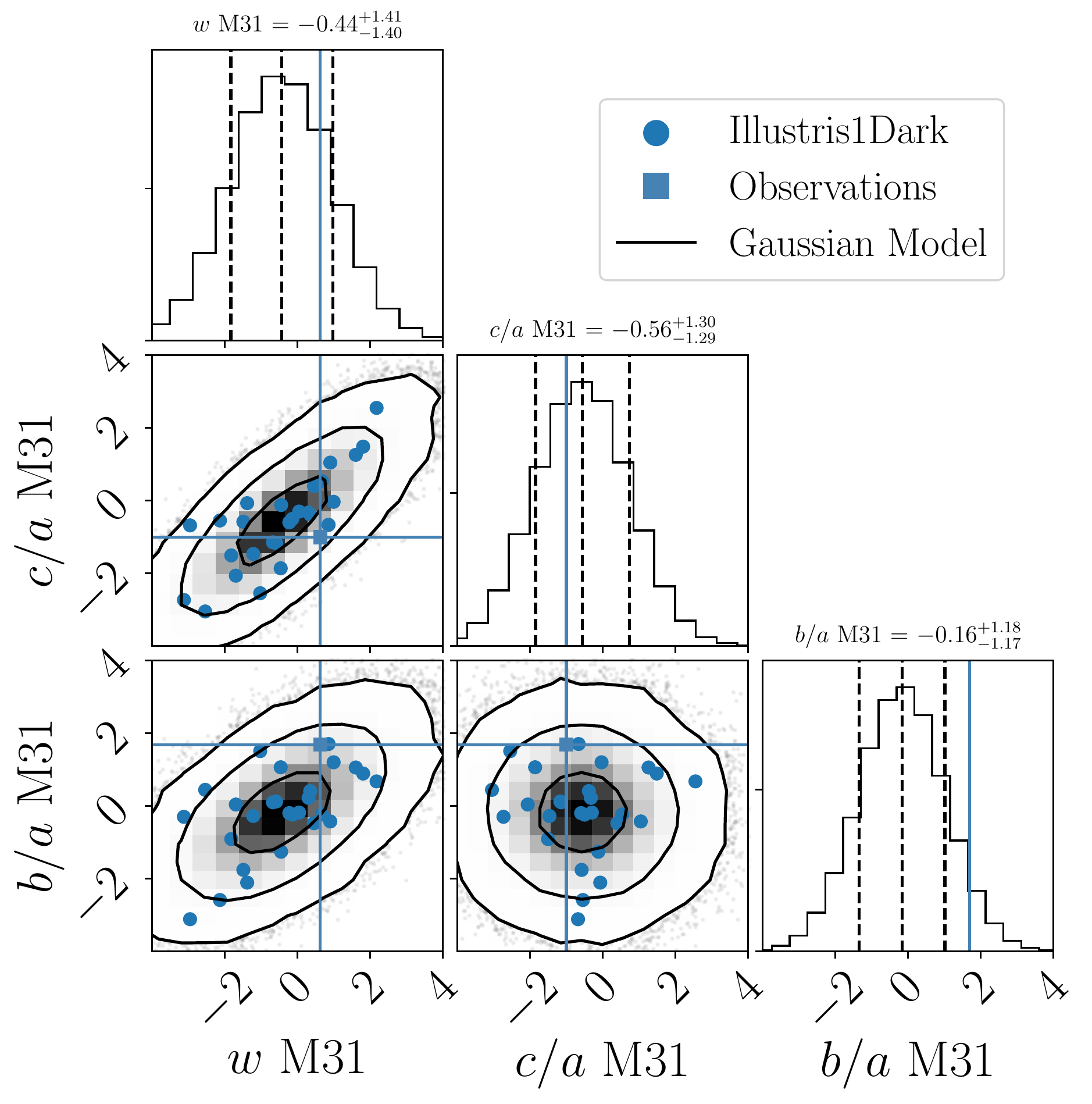}
\includegraphics[width=0.45\textwidth]{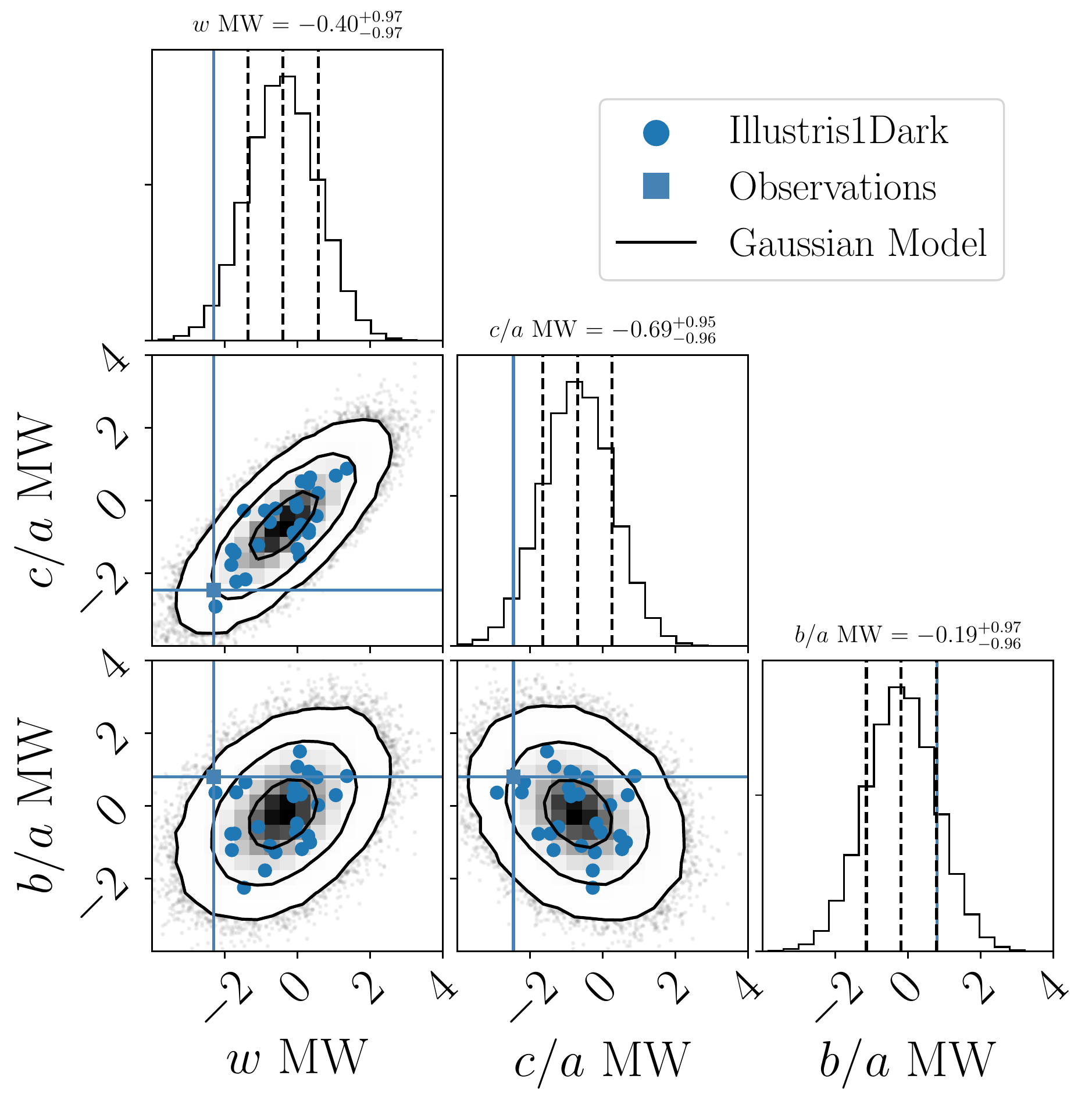}
\caption{
Same layout as Figure \ref{fig:normalized_illustris1dark}. 
This time the distribution comes from the multivariate gaussian model 
with its parameters estimated from the Illustris-1-Dark results.
The corresponding plots for Illustris-1 and ELVIS are
in Figures \ref{fig:all_plots_illustris1} and \ref{fig:all_plots_elvis},
respectively. \label{fig:gaussian_illustris1dark}}     
\end{figure*}

\section{Results}
\label{sec:results}

Figure \ref{fig:normalized_n} shows, as a function of the number of
satellites, the normalized plane width $w$, $c/a$ ratio and
$b/a$ ratio from observations. 
The most oustanding feature of this plot is that for $w$ and $c/a$
ratio the MW is always two standard deviations away from the mean,
while M31 is always within 2 standard deviations.
The results for the $b/a$ ratio are always 2 standard deviations both
for the MW and M31. 
This confirms the extreme planar distribution for the MW and the more
spherical distribution for the M31.

In the following subsections we describe in detail the results from
Illustris-1-Dark for $w$, $c/a$ and $b/a$ for a fixed number of
satellites $N_s=11$.
The results from the other two simulations are included in the
Appendix \ref{appendix:plots}. 
The full covariance matrices are in the Appendix
\ref{appendix:matrices}.

\subsection{Physical Units}

Figure \ref{fig:physical_illustris1dark} summarizes the results in
physical units.
It compares three different groups: observations, simulations and the
observations randomized.
The left panel shows the results for M31 and the panel on the right for
the MW.

The most interesting outcome is that the MW plane width is smaller
than $\approx 98\%$ of the planes computed from the randomized distribution;
its  $c/a$ ratio is also significantly lower than the measured values for
the randomized distributions.   
On the other hand the results for $M31$ for the same quantities are
within $1\sigma$ of the randomized results.

The comparison against observations also provides an extreme picture
for the MW. 
In this case the mean witdh from simulation is almost twice as the
mean width from the randomization. 
A possible criticism to this result is that the typical halo mass in
the simulation is larger than the it is for the MW. 
For this reason we renormalize all quantities ($w$, $c/a$ and $b/a$)
to the mean value and standard deviation provided by its spherical
randomization. 
In this way the physical scale of the halo becomes a second order
effect and only the deviations from asphericity are protagonist.

\subsection{Normalized Units}

Figure \ref{fig:normalized_illustris1dark} summarizes  the results of
renormalizing to the randomized results.
The layout is the same as Figure \ref{fig:physical_illustris1dark}.
By construction, the relative position of the observations and the
randomized results keeps constant; The MW continues to be atypical
with respect to its own randomized distribution.

The most notorious difference comes from the changes in the
simulations. 
The previous large difference betwen the width distribution from
MW observations and simulations is reduced. 
The reason is that normalized quantities are not dependent on the
physical scale anymore; only the deviations from asphericity are
important.

The distributions for the normalized quantities from simulations are
well described by gaussians. 
The parameters of the gaussian model (covariance matrix and mean)
are different depending on the simulation and number of particles, but
all of them are well described by the multivariate gaussian.

\subsection{Multivariate Gaussian model}

Figure \ref{fig:gaussian_illustris1dark} summarized the results from
 computing the covariance matrix and mean vector in
 Eq.\ref{eq:multivariate} from the normalized quantities obtained from
 the Illustris-1-Dark simulation.
The distributions in this Figure are computed from $10^6$ points
generated with the multivariate gaussian. 

This nicely summarizes the results we had in the previous
sections. 
The left hand triangular plot shows how M31 falls within the $2\sigma$
contours in the 2D scatter plots;
while the right hand plot clearly places the MW observations at the
border of the $3\sigma$ range in the joint distributions that involve
the width $w$.  

In both cases the strongest positive correlation is present for the
width and the $c/a$ axis ratio. 
A weaker correlation is present for the width and the $b/a$ axis
ratio.  
The weakest correlation, positive or negative depending on the
simulation, is present for the $c/a$ ratio and $b/a$ ratio.

The mean values also hold valuable information. 
For the three quantities the mean is negative, a quantification trend
that the satellite distribution from LCDM simulations tend to be aspherical.
This trend has a very few exceptions. 
Nmely the ELVIS results for M31 with $N_s=11,12$ and the Illustris-1
results for the $MW$ with $N_s=11$, which are consistent with an
spherical distribution. 
We also computed the correlation between the M31 and MW results,
finding that it is very weak and can be safely discarded. 

The explicity formulation for this probability distribution allows us
to generate large samples with asphericity properties consistent with
the parent simulation. 
From this samples we can explicitly compute the fraction of systems
that, for instance, show the extreme MW values.

\subsection{Number of Expected LG Systems}

\begin{figure*}
\centering
\includegraphics[width=0.32\textwidth]{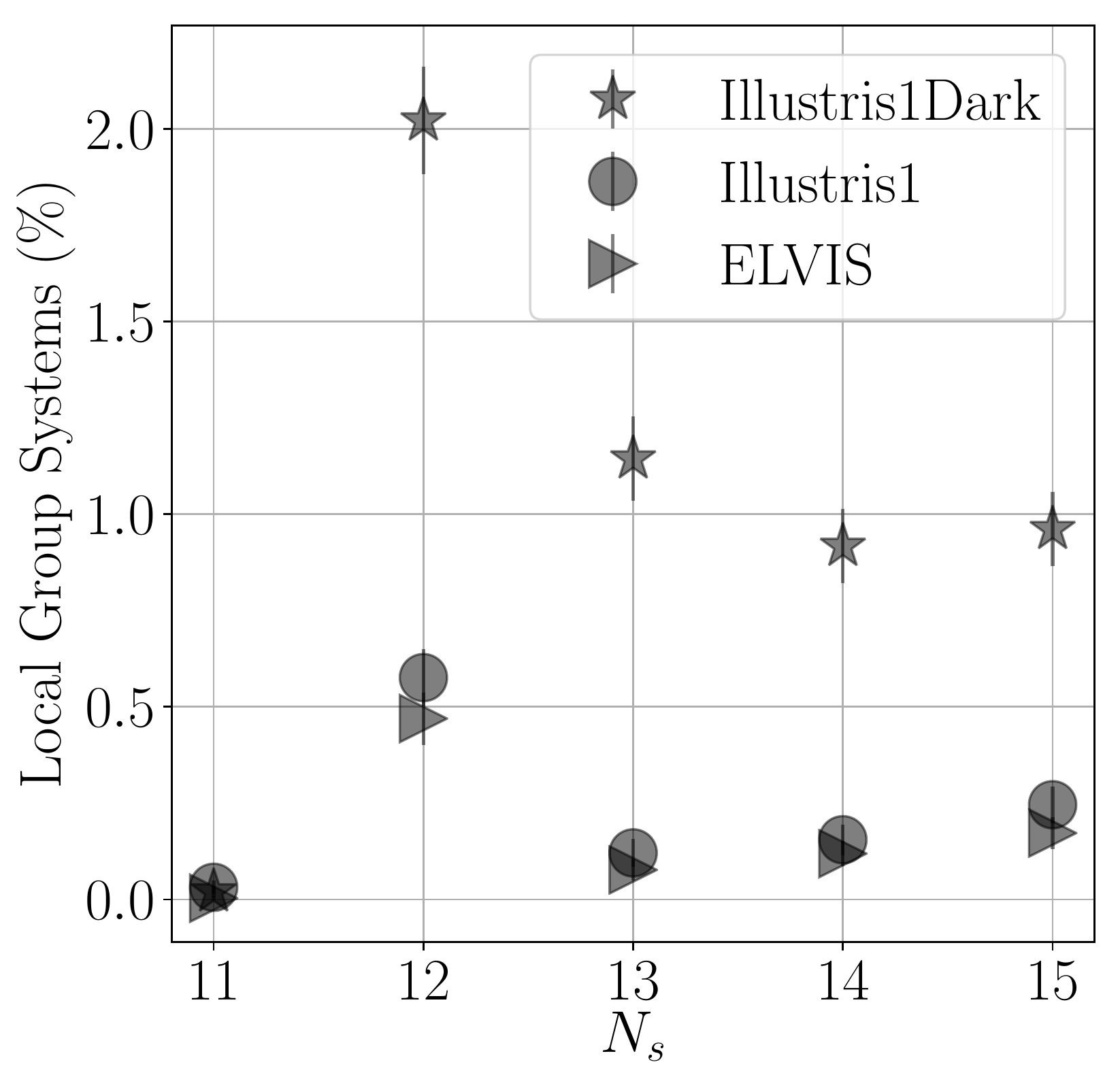}
\includegraphics[width=0.32\textwidth]{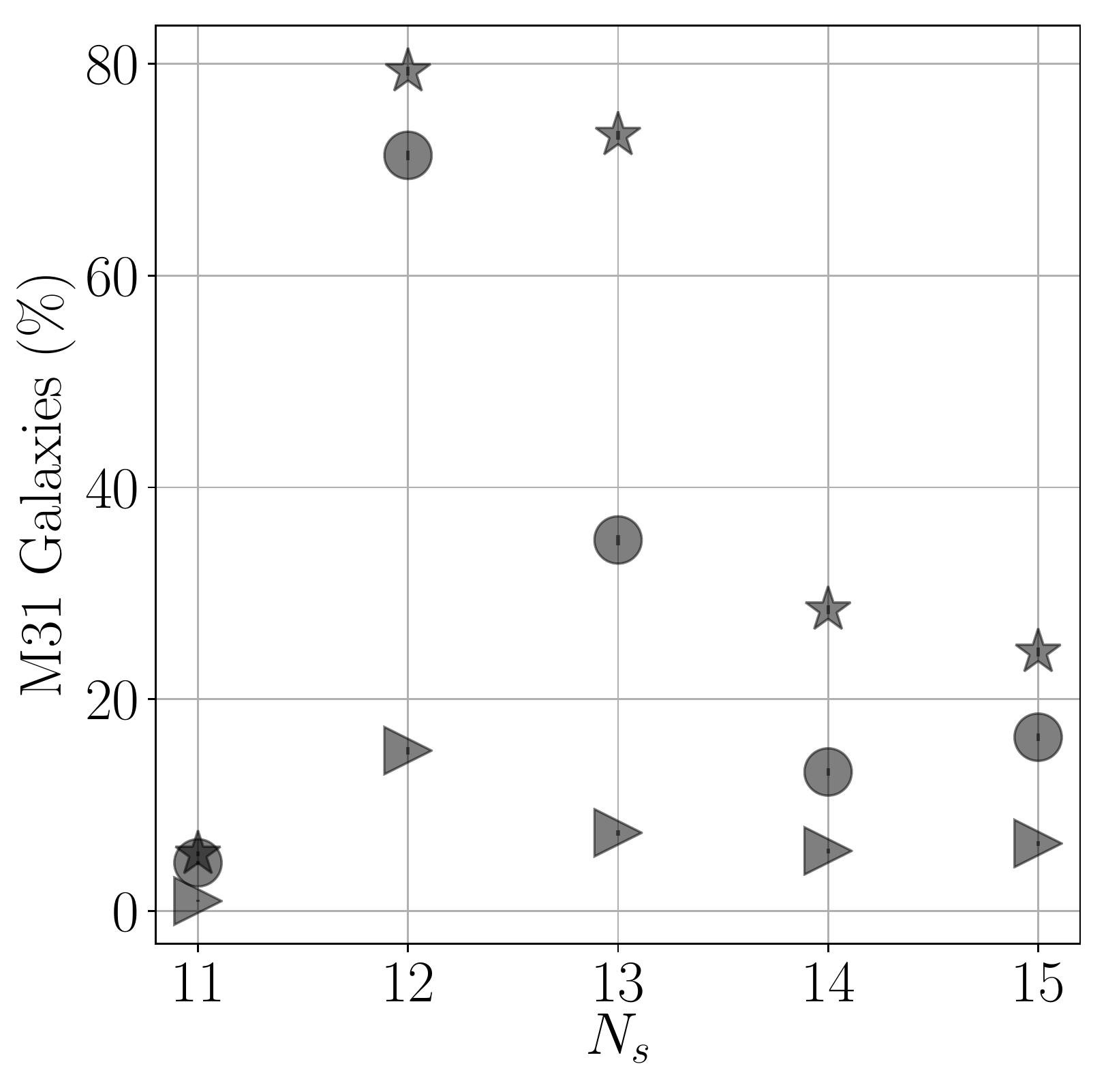}
\includegraphics[width=0.32\textwidth]{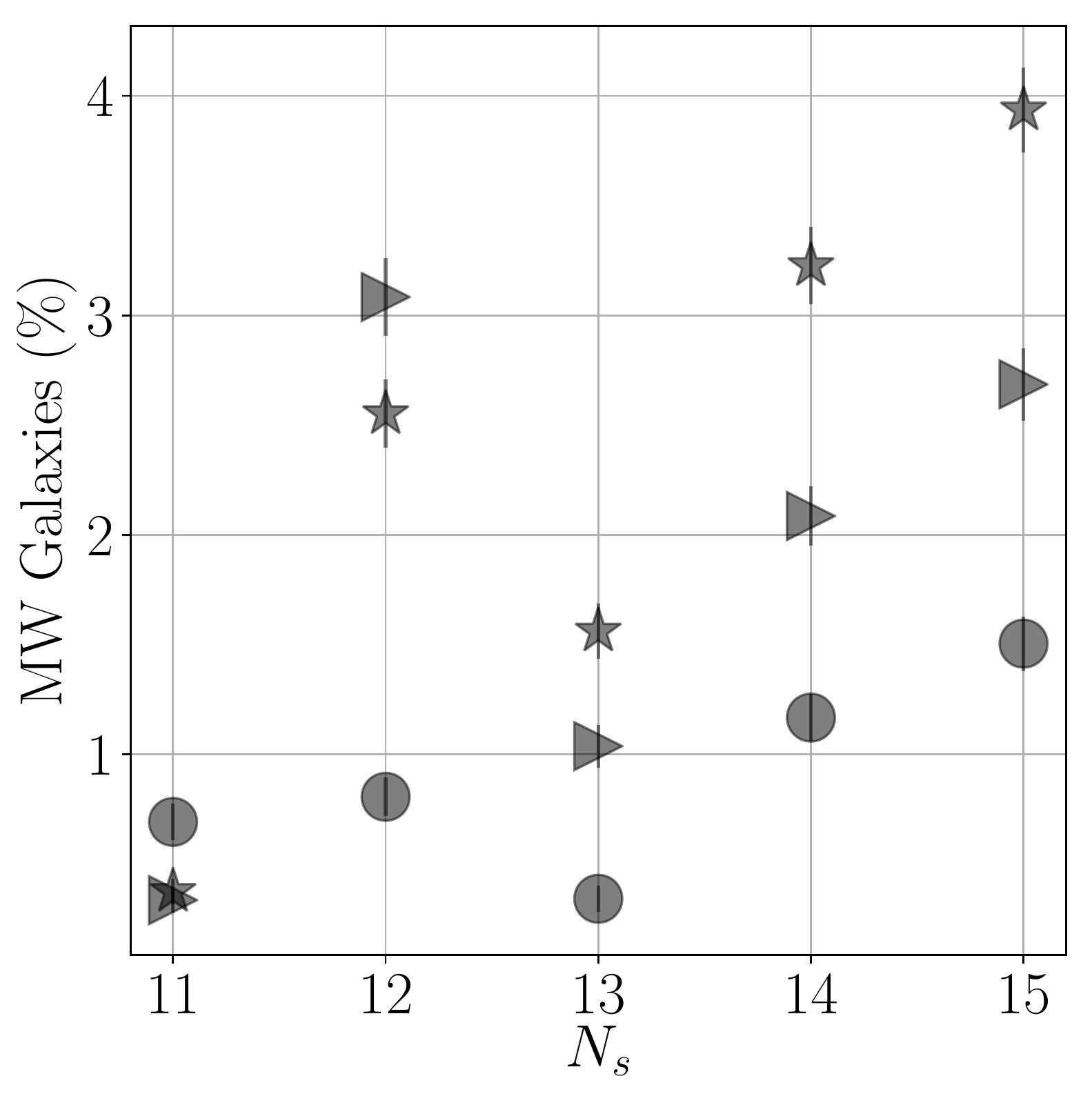}
\caption{
Percentage of systems that have asphericity scalars as extreme as the
LG (left), M31 (middle) and MW (right).  
This percentage is computed as a function of satellite number for all
the three simulations (Illustris-1-Dark, Illustris-1 and ELVIS).
The common trend is that at most of $2\%$ of LG-like pairs are expected to
be as extreme as observations, with most of the weight of this
atypicality falls onto the MW; at most $4\%$ of similar systems have
such an aspherical satellite distribution. 
\label{fig:expected_number}}
\end{figure*}

Using the multivariate gaussian model we generate $10^3$ samples, each
sample containing $10^4$ pairs, where each pair member is drawn from
the corresponding multivariate gaussian distribution. 

We consider that a sampled system is similar to the M31/MW galaxy if the
absolute distance of each of its normalized characteristics ($w$, $c/a$, $b/a$) to the
sample mean is equal or larger than the absolute distance of the observational
values to the sample mean.   
That is, we perform a double-tailed test around the mean using the
observational values as a threshold.  

Figure \ref{fig:expected_number} summarizes the results from this
experiment as a function of satellite number for all three simulations:
Illustris-1-Dark, Illustris-1 and ELVIS.

Considering the joint distribution of M31 and MW we find that at most
$2\%$ of the pairs are expected to be similar to the LG.
In a three dimensional gaussian distribution, having a $1\sigma$,
$2\sigma$ and $3\sigma$ interval corresponds to having respectively $19 \%$, $73 \%$ and
$97 \%$ of the total of points in the distribution.
With this result in mind the $LG$ has the same degree of atypicality
as a $3\sigma$ outlier. 

For M31, between $5\%$ and $80\%$ of the pairs have a satellite
distribution as aspherical as the one observed in M31.
This fraction drops
dramatically for the MW where only $0.02\%$ to $4\%$ of the satellites
are expected to have as extreme aspherical distributions as the MW.
A consistent finding is that the M31 fraction is always larger than
the MW fraction at fixed $N_s$ and simulation.

Among the three simulations, the results inferred from ELVIS data show
the lowest fraction of M31 and MW systems; for Illustris-1-Dark we have
the highest fraction of M31/MW systems. 
The results from Illustris-1 are in between these two simulations, but
closer to ELVIS. 

A probable reason for these trends is the different median mass
for the MW/M31 halos in the pairs from these simulations. 
For instance for the MW halo the median maximum circular velocity is
$\sim 160$ \kms, $\sim 150$ \kms and $\sim 120$ \kms in the ELVIS,
Illustris-1 and Illustris-1-Dark simulations, respectively.
For the M31 halo this median velocity is $\sim 200$\kms both for the
ELVIS and Illustris1 simulations, while for the Illustris-1-Dark it is
$\sim 160$ \kms. 

Another element that influences this trend is the simulated physics.
This is evident in the comparison between Illustris-1 and Illustris-1-Dark.
These two simulations share the same characteristics except for the
inclusion of baryonic physics in Illustris-1.
We find that extreme MW systems are easier to find in the DM only
simulation. 
The results listed in Appendix \ref{appendix:matrices} show that
halos are closer to spherical once baryonic effects are included. 
For instance for $N_p=11$ the mean value for the normalized width goes
closer to zero, from $-0.43\pm 0.04$  to $-0.03\pm0.06$, and the same
is true for the $c/a$ ratio that changes from $-0.56\pm0.06$ in the DM
only simulation to $-0.24\pm 0.06$ in Illustris-1.  
This trend has been reported before for the DM distribution in the
whole halo \cite{2013MNRAS.429.3316B}.
Our results suggest that it also hold for the satellite distribution.
We postpone a detailed quantification of this effect for a future
study.

\subsection{How to understand an atypical MW}

The highly aspherical satellite distribution in the MW is another piece of
information that points at an atypical configuration in LCDM.
We also have the number of satellites as bright as the Magellanic
Clouds, only expected in $5\%$ of galaxies
\citep{2011ApJ...743..117B} and 
the satellite velocities around the MW with a radial/tangential
anisotropy only expected in $3\%$ of systems in LCDM
\citep{2017MNRAS.468L..41C}. 
One could also add the atypical kinematics of M31 with a very low
tangential velocity, which is only expected in less than $1\%$ of the pairs
with similar environmental characteristics \citep{ForeroRomero2013}.  
With the small numbers that we have in the simulations we used we
cannot study properly those correlations. 
We cannot even find a clear trend linking asphericity to halo mass or
another pair property.

Assuming that these atypical properties are independent one would need at
least $10^6$ pairs in order to find a single pair that meets all four
characteristics (aspherical satellite distribution, bright Magellanic
Clouds, satellite velocity anisotropy and atypical pair kinematics).
Using the number density for pairs in Illustris-1 ($2\times 10^{-5}$
pairs/Mpc$^{3}$) this would imply a cubic simulation volume of $5$ Gpc on a
side, something unfeasible under current technology.
However, studying two characteristics at a time to find possible
correlations, and at least one pair resembling observations, reduces
the box size to $500$ Mpc, something that is close to current
simulations such as Illustris-TNG \citep{2018MNRAS.473.4077P}.

\section{Conclusions}\label{sec:conclusions}

In this paper we developed and demonstrated a method to quantify the
asphericity of the satellite distribution around the MW and M31.  
In the interest of keeping the demonstration straightforward and robust we
focus on the asphericity estimates for a fixed number of the brightest
satellites around each galaxy. 

The method uses as a reference the spherically randomized data
of the system under study \citep{2017AN....338..854P}.  
To this end, we first measure the width and axis ratios for the satellite
distributions of interest. 
Then, we measure the same quantities for the same set of points after
the spherical randomization process.
Finally, we renormalize the initial results to the mean value and
standard deviation computed from the randomized data.  

We found that these normalized quantities are well described by
a multivariate gaussian distribution 
We estimated the mean and covariance of these distributions using
the results of LG pairs coming from three different numerical
simulations (Illustris-1, Illustris-1-Dark and ELVIS). 
Finally, we compared the observational results against the
distributions derived from the simulations

We quantified that in the best case (Illustris-1-Dark) the degree of
asphericity in the observed LG is only expected in $2\%$ of the
pairs. 
This places the LG as a $3\sigma$ outlier.   
The weight to explain this atypical result is not distributed equally
between the MW and M31. 
While M31 presents a fully typical asphericity in the expectations
from LCDM, the MW shows aspherical deviations in plane width and the
major-to-minor axis ratio highly atypical in the framework of LCDM,
confirming the original hint by \cite{2005A&A...431..517K} and more
recent results \citep{2012MNRAS.423.1109P,2015ApJ...815...19P}. 
We estimated that with the M31 between $5\%$ and $80\%$ of the pairs
show aspherical characteristics larger than M31, while this fraction
drops to less than $4\%$ for the MW.  
These fractions are robust to changes in the numerical simulations,
the criteria used to define the pairs, the way to select the
satellites and to the methods used to estimate the parameters of the
multivariate normal distribution.

The focus of our approach was building an analytic probability
distribution for the observables of interest, instead of trying to
find simulated objects that fulfill different observational criteria. 
This approach is particularly useful in the case of atypical
observables, as it allows for an atypicality quantification without
building explicit samples of objects that are already scarce and
difficult to find in simulations.

An extension of this framework to outliers in higher order deviations
(i.e. coherent \emph{velocity} structures) should also be possible, 
provided that an analytic probability distribution for the scalars of
interest can be built.  
The approach presented here is also useful to gauge the influence of 
influence of different physical elements includes the simulation. 
For instance, in our case the data hints towards rounder satellite
distributions in simulations that include baryonic effects.  

The LG atypicality should be seen as an opportunity to constrain in
great detail the environment that allowed such a pattern to emerge. 
Although broad correlations between LG assembly, pair kinematics, halo
shapes and satellite distributions are expected in LCDM
\citep{2011MNRAS.417.1434F,2014MNRAS.443.1090F,2015ApJ...799...45F,2015MNRAS.452.1052L},
we claim that detailed studies of the satellite asphericity as a
function of halo mass and cosmic web environment are still missing to
understand what features in the initial conditions of our LG are
responsible for the extreme features observed in its satellite
distribution. 

\section*{Acknowledgments} 
JEFR acknowledges support from COLCIENCIAS Contract No. 287-2016,
Project 1204-712-50459 and the European Union's Horizon 2020 Research and Innovation
Programme under the Marie Sk\l{}odowska-Curie grant agreement No 73437
(LACEGAL).   
We thank the referee, Pavel Kroupa and Marcel Pawlowski for useful
comments that improved the clarity of the paper.
  
\bibliographystyle{mnras}

\appendix

\section{Physical Characteristics of the Isolated Pairs samples}
\label{appendix:physical}

\begin{figure}
\centering
\includegraphics[width=0.30\textwidth]{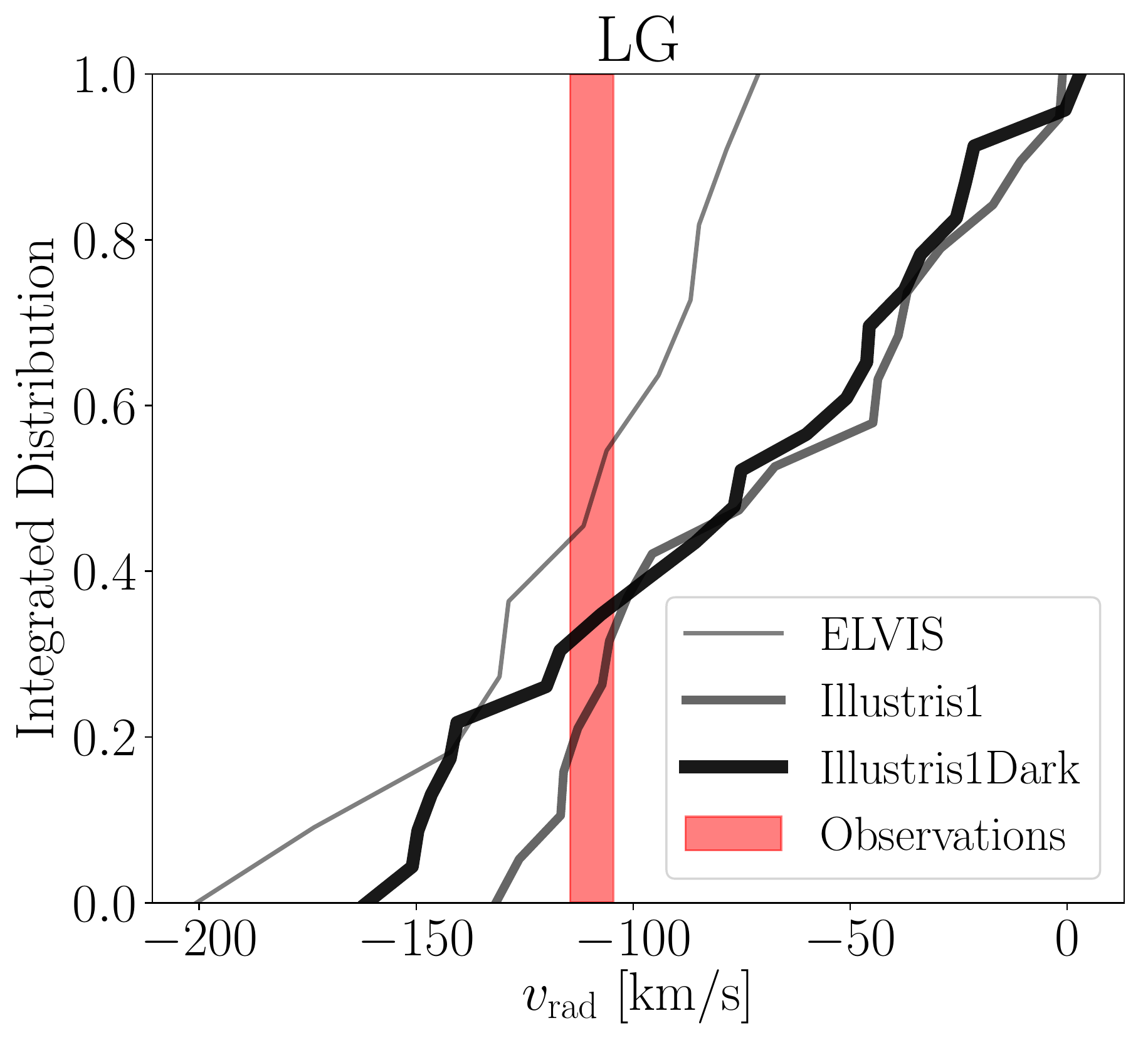}
\includegraphics[width=0.30\textwidth]{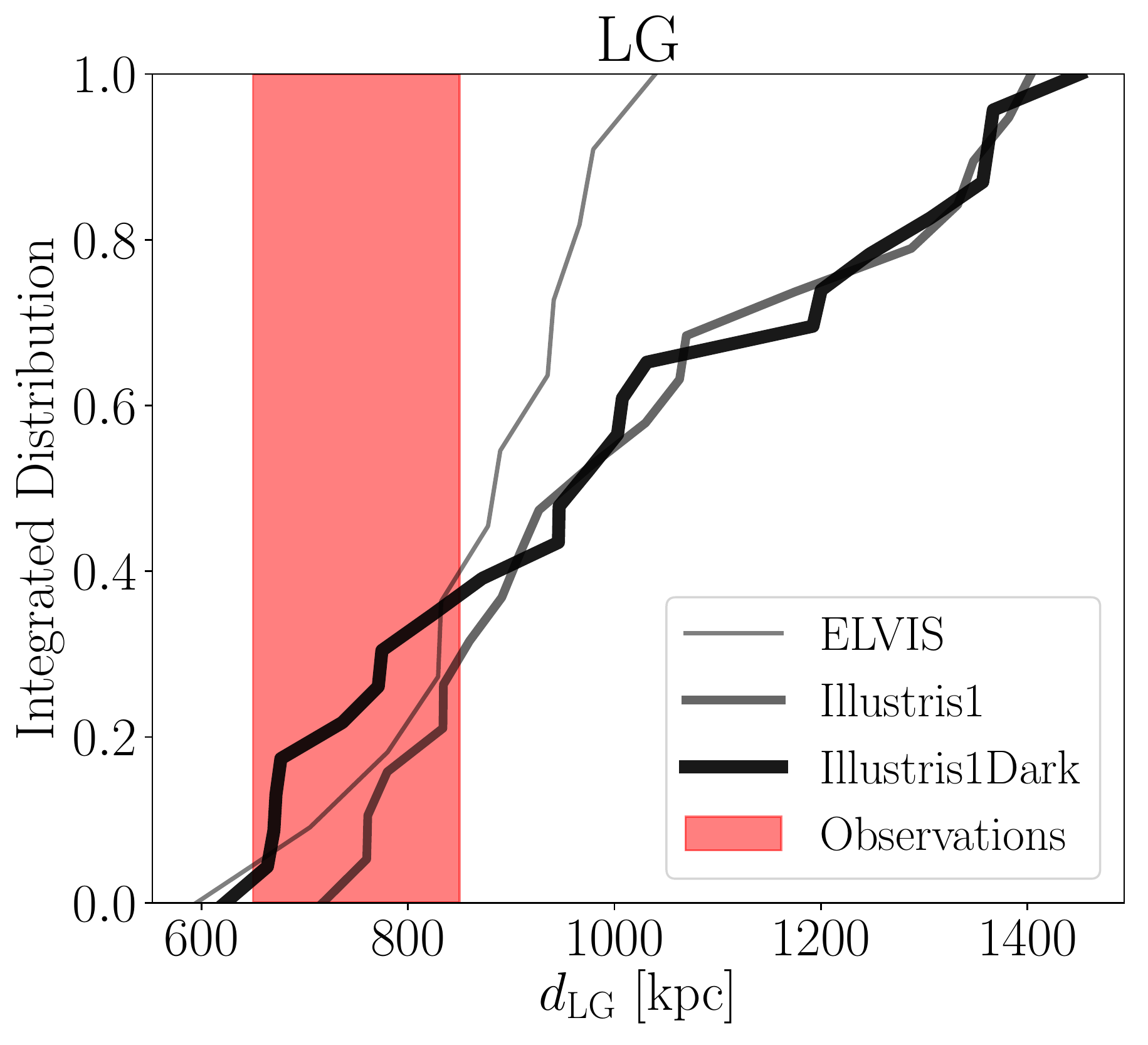}
\includegraphics[width=0.30\textwidth]{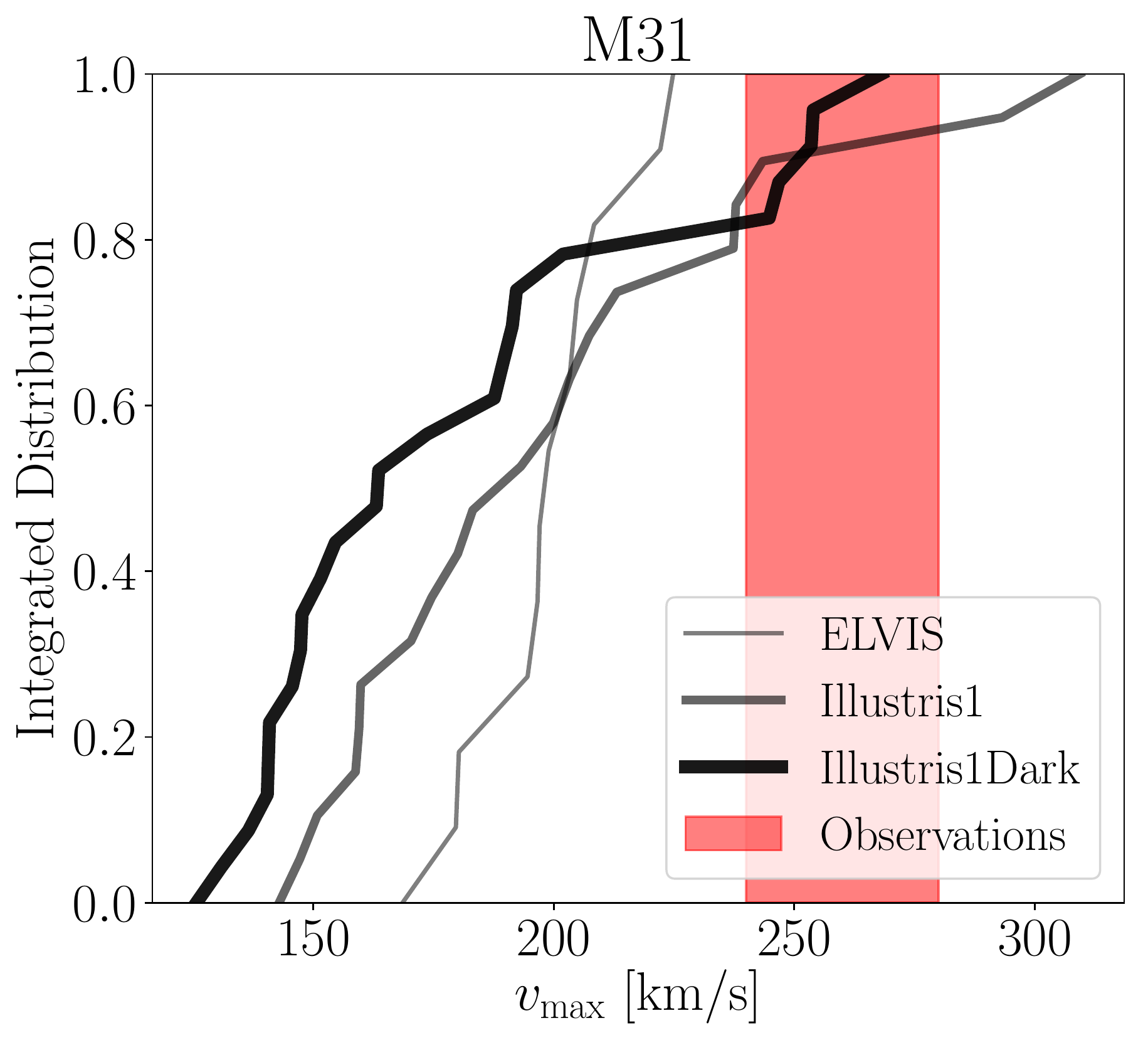}
\includegraphics[width=0.30\textwidth]{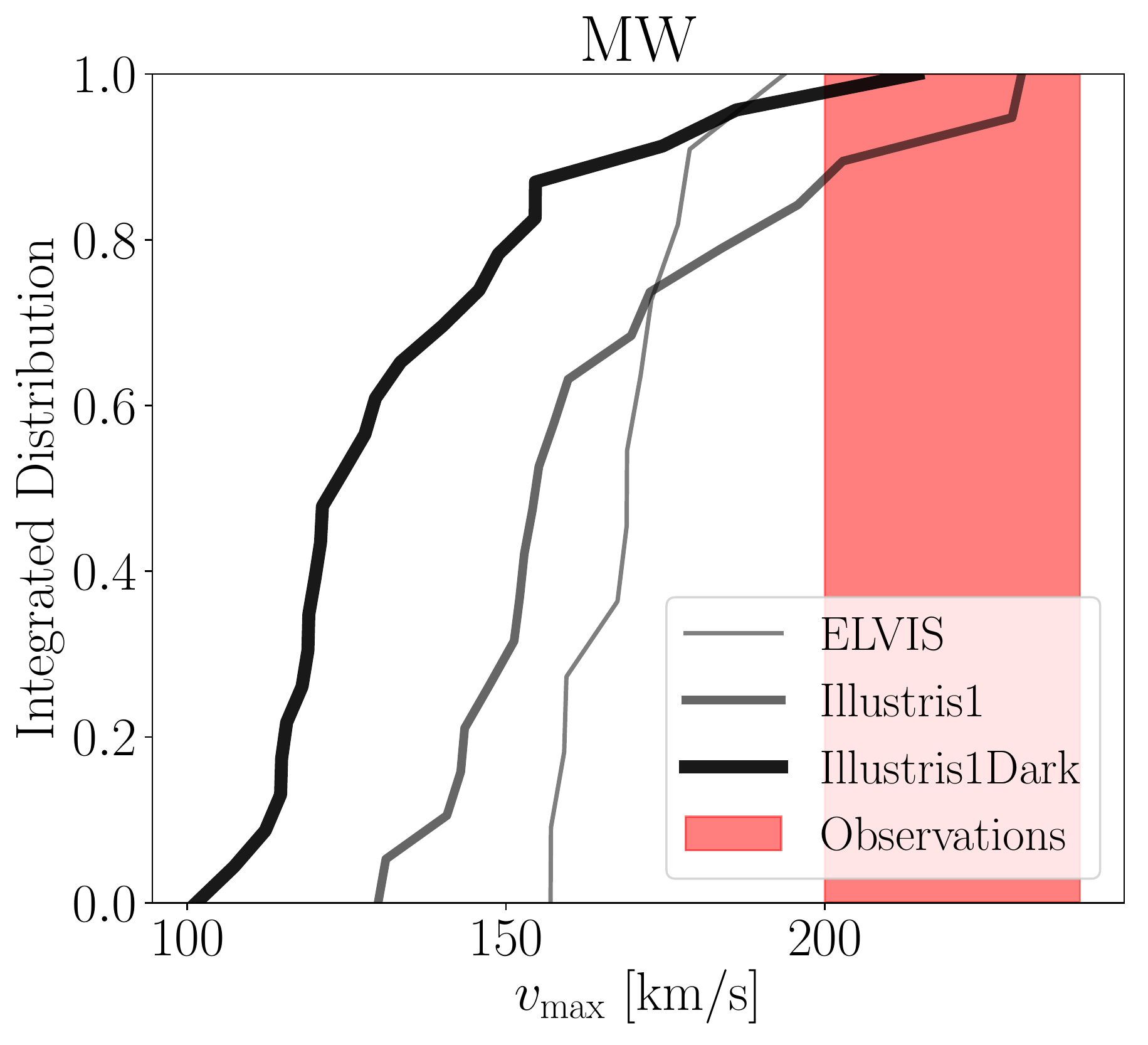}
\caption{Physical characteristics of the LG pairs selected in the
  simulations. All plots show the integrated distributions. The
  physical properties are the radial comoving velocity between the MW
  and M31, the radial separation between the MW and M31 and the
  maximum circular velocity for the M31 and MW dark matter halo.
\label{fig:physical_pairs}}
\end{figure}

\section{Covariance Matrices and Mean value vectors}
\label{appendix:matrices}

\subsection{Illustris-1-Dark, M31, $N_s=11$}
\[
\Sigma=
\begin{bmatrix}
2.02 \pm 0.09 & 1.46 \pm 0.09 & 0.97 \pm 0.07\\
1.46 \pm 0.09 & 1.69 \pm 0.10 & -0.01 \pm 0.05\\
0.97 \pm 0.07 & -0.01 \pm 0.05 & 1.42 \pm 0.08\\
\end{bmatrix}
\]
\[
\mu=
\begin{bmatrix}
-0.43 \pm 0.06 & -0.56 \pm 0.06 & -0.17 \pm 0.05\\
\end{bmatrix}
\]
\subsection{Illustris-1-Dark, MW, $N_s=11$}
\[
\Sigma=
\begin{bmatrix}
0.94 \pm 0.04 & 0.71 \pm 0.04 & 0.36 \pm 0.03\\
0.71 \pm 0.04 & 0.93 \pm 0.05 & -0.27 \pm 0.03\\
0.36 \pm 0.03 & -0.27 \pm 0.03 & 0.94 \pm 0.04\\
\end{bmatrix}
\]
\[
\mu=
\begin{bmatrix}
-0.40 \pm 0.04 & -0.69 \pm 0.04 & -0.19 \pm 0.04\\
\end{bmatrix}
\]

\subsection{Illustris-1, M31, $N_s=11$}
\[
\Sigma=
\begin{bmatrix}
1.59 \pm 0.08 & 1.23 \pm 0.07 & 0.80 \pm 0.06\\
1.23 \pm 0.07 & 1.44 \pm 0.08 & 0.01 \pm 0.06\\
0.80 \pm 0.06 & 0.01 \pm 0.06 & 1.26 \pm 0.08\\
\end{bmatrix}
\]
\[
\mu=
\begin{bmatrix}
0.03 \pm 0.06 & -0.24 \pm 0.06 & -0.01 \pm 0.05\\
\end{bmatrix}
\]
\subsection{Illustris-1, MW, $N_s=11$}
\[
\Sigma=
\begin{bmatrix}
1.01 \pm 0.05 & 0.72 \pm 0.05 & 0.62 \pm 0.04\\
0.72 \pm 0.05 & 0.78 \pm 0.05 & 0.12 \pm 0.03\\
0.62 \pm 0.04 & 0.12 \pm 0.03 & 0.80 \pm 0.04\\
\end{bmatrix}
\]
\[
\mu=
\begin{bmatrix}
-0.07 \pm 0.05 & -0.45 \pm 0.04 & 0.02 \pm 0.04\\
\end{bmatrix}
\]

\subsection{ELVIS, M31, $N_s=11$}
\[
\Sigma=
\begin{bmatrix}
1.36 \pm 0.22 & 1.59 \pm 0.26 & 0.24 \pm 0.07\\
1.59 \pm 0.26 & 2.00 \pm 0.31 & 0.06 \pm 0.08\\
0.24 \pm 0.07 & 0.06 \pm 0.08 & 0.52 \pm 0.03\\
\end{bmatrix}
\]
\[
\mu=
\begin{bmatrix}
0.45 \pm 0.08 & 0.22 \pm 0.10 & 0.07 \pm 0.05\\
\end{bmatrix}
\]
\subsection{ELVIS, MW, $N_s=11$}
\[
\Sigma=
\begin{bmatrix}
1.53 \pm 0.16 & 0.91 \pm 0.17 & 0.98 \pm 0.09\\
0.91 \pm 0.17 & 0.89 \pm 0.16 & 0.18 \pm 0.06\\
0.98 \pm 0.09 & 0.18 \pm 0.06 & 1.19 \pm 0.13\\
\end{bmatrix}
\]
\[
\mu=
\begin{bmatrix}
-0.46 \pm 0.08 & -0.28 \pm 0.06 & -0.77 \pm 0.07\\
\end{bmatrix}
\]

\subsection{Illustris-1-Dark, M31, $N_s=12$}
\[
\Sigma=
\begin{bmatrix}
2.34 \pm 0.11 & 1.72 \pm 0.11 & 1.10 \pm 0.06\\
1.72 \pm 0.11 & 1.72 \pm 0.11 & 0.35 \pm 0.05\\
1.10 \pm 0.06 & 0.35 \pm 0.05 & 1.21 \pm 0.07\\
\end{bmatrix}
\]
\[
\mu=
\begin{bmatrix}
-0.58 \pm 0.07 & -0.49 \pm 0.06 & -0.41 \pm 0.05\\
\end{bmatrix}
\]
\subsection{Illustris-1-Dark, MW, $N_s=12$}
\[
\Sigma=
\begin{bmatrix}
1.25 \pm 0.06 & 0.97 \pm 0.06 & 0.62 \pm 0.04\\
0.97 \pm 0.06 & 1.19 \pm 0.06 & -0.11 \pm 0.05\\
0.62 \pm 0.04 & -0.11 \pm 0.05 & 1.19 \pm 0.04\\
\end{bmatrix}
\]
\[
\mu=
\begin{bmatrix}
-0.38 \pm 0.05 & -0.55 \pm 0.05 & -0.28 \pm 0.05\\
\end{bmatrix}
\]

\subsection{Illustris-1, M31, $N_s=12$}
\[
\Sigma=
\begin{bmatrix}
1.01 \pm 0.06 & 0.74 \pm 0.04 & 0.59 \pm 0.06\\
0.74 \pm 0.04 & 1.03 \pm 0.05 & -0.16 \pm 0.04\\
0.59 \pm 0.06 & -0.16 \pm 0.04 & 1.12 \pm 0.06\\
\end{bmatrix}
\]
\[
\mu=
\begin{bmatrix}
-0.09 \pm 0.05 & -0.23 \pm 0.05 & -0.33 \pm 0.05\\
\end{bmatrix}
\]
\subsection{Illustris-1, MW, $N_s=12$}
\[
\Sigma=
\begin{bmatrix}
1.00 \pm 0.05 & 0.69 \pm 0.05 & 0.64 \pm 0.05\\
0.69 \pm 0.05 & 0.87 \pm 0.06 & -0.06 \pm 0.04\\
0.64 \pm 0.05 & -0.06 \pm 0.04 & 1.06 \pm 0.06\\
\end{bmatrix}
\]
\[
\mu=
\begin{bmatrix}
-0.16 \pm 0.05 & -0.46 \pm 0.04 & -0.12 \pm 0.05\\
\end{bmatrix}
\]

\subsection{ELVIS, M31, $N_s=12$}
\[
\Sigma=
\begin{bmatrix}
1.17 \pm 0.09 & 1.60 \pm 0.14 & 0.26 \pm 0.05\\
1.60 \pm 0.14 & 2.37 \pm 0.21 & 0.11 \pm 0.08\\
0.26 \pm 0.05 & 0.11 \pm 0.08 & 0.44 \pm 0.04\\
\end{bmatrix}
\]
\[
\mu=
\begin{bmatrix}
0.66 \pm 0.07 & 0.49 \pm 0.10 & 0.18 \pm 0.04\\
\end{bmatrix}
\]
\subsection{ELVIS, MW, $N_s=12$}
\[
\Sigma=
\begin{bmatrix}
2.00 \pm 0.18 & 1.14 \pm 0.17 & 1.31 \pm 0.11\\
1.14 \pm 0.17 & 0.98 \pm 0.17 & 0.39 \pm 0.05\\
1.31 \pm 0.11 & 0.39 \pm 0.05 & 1.34 \pm 0.16\\
\end{bmatrix}
\]
\[
\mu=
\begin{bmatrix}
-0.50 \pm 0.10 & -0.34 \pm 0.07 & -0.67 \pm 0.08\\
\end{bmatrix}
\]

\subsection{Illustris-1-Dark, M31, $N_s=13$}
\[
\Sigma=
\begin{bmatrix}
1.79 \pm 0.09 & 1.19 \pm 0.08 & 0.90 \pm 0.06\\
1.19 \pm 0.08 & 1.12 \pm 0.07 & 0.17 \pm 0.03\\
0.90 \pm 0.06 & 0.17 \pm 0.03 & 1.12 \pm 0.07\\
\end{bmatrix}
\]
\[
\mu=
\begin{bmatrix}
-0.61 \pm 0.06 & -0.67 \pm 0.05 & -0.37 \pm 0.05\\
\end{bmatrix}
\]
\subsection{Illustris-1-Dark, MW, $N_s=13$}
\[
\Sigma=
\begin{bmatrix}
1.30 \pm 0.07 & 1.09 \pm 0.06 & 0.52 \pm 0.05\\
1.09 \pm 0.06 & 1.42 \pm 0.08 & -0.18 \pm 0.06\\
0.52 \pm 0.05 & -0.18 \pm 0.06 & 1.09 \pm 0.04\\
\end{bmatrix}
\]
\[
\mu=
\begin{bmatrix}
-0.38 \pm 0.05 & -0.54 \pm 0.05 & -0.30 \pm 0.05\\
\end{bmatrix}
\]

\subsection{Illustris-1, M31, $N_s=13$}
\[
\Sigma=
\begin{bmatrix}
1.36 \pm 0.09 & 0.89 \pm 0.05 & 0.97 \pm 0.06\\
0.89 \pm 0.05 & 0.92 \pm 0.05 & 0.22 \pm 0.05\\
0.97 \pm 0.06 & 0.22 \pm 0.05 & 1.28 \pm 0.06\\
\end{bmatrix}
\]
\[
\mu=
\begin{bmatrix}
-0.07 \pm 0.06 & -0.17 \pm 0.05 & -0.30 \pm 0.06\\
\end{bmatrix}
\]
\subsection{Illustris-1, MW, $N_s=13$}
\[
\Sigma=
\begin{bmatrix}
0.95 \pm 0.08 & 0.90 \pm 0.09 & 0.33 \pm 0.05\\
0.90 \pm 0.09 & 1.29 \pm 0.10 & -0.24 \pm 0.05\\
0.33 \pm 0.05 & -0.24 \pm 0.05 & 0.89 \pm 0.08\\
\end{bmatrix}
\]
\[
\mu=
\begin{bmatrix}
-0.13 \pm 0.05 & -0.47 \pm 0.06 & -0.09 \pm 0.05\\
\end{bmatrix}
\]

\subsection{ELVIS, M31, $N_s=13$}
\[
\Sigma=
\begin{bmatrix}
1.20 \pm 0.11 & 1.47 \pm 0.16 & 0.48 \pm 0.09\\
1.47 \pm 0.16 & 2.02 \pm 0.24 & 0.35 \pm 0.08\\
0.48 \pm 0.09 & 0.35 \pm 0.08 & 0.56 \pm 0.10\\
\end{bmatrix}
\]
\[
\mu=
\begin{bmatrix}
0.77 \pm 0.07 & 0.61 \pm 0.10 & 0.27 \pm 0.05\\
\end{bmatrix}
\]
\subsection{ELVIS, MW, $N_s=13$}
\[
\Sigma=
\begin{bmatrix}
1.82 \pm 0.19 & 0.99 \pm 0.15 & 1.43 \pm 0.19\\
0.99 \pm 0.15 & 0.92 \pm 0.14 & 0.39 \pm 0.08\\
1.43 \pm 0.19 & 0.39 \pm 0.08 & 1.56 \pm 0.23\\
\end{bmatrix}
\]
\[
\mu=
\begin{bmatrix}
-0.26 \pm 0.09 & -0.32 \pm 0.06 & -0.37 \pm 0.08\\
\end{bmatrix}
\]

\subsection{Illustris-1-Dark, M31, $N_s=14$}
\[
\Sigma=
\begin{bmatrix}
1.60 \pm 0.11 & 1.34 \pm 0.09 & 0.68 \pm 0.05\\
1.34 \pm 0.09 & 1.32 \pm 0.08 & 0.28 \pm 0.04\\
0.68 \pm 0.05 & 0.28 \pm 0.04 & 0.85 \pm 0.05\\
\end{bmatrix}
\]
\[
\mu=
\begin{bmatrix}
-0.33 \pm 0.06 & -0.53 \pm 0.05 & -0.15 \pm 0.04\\
\end{bmatrix}
\]
\subsection{Illustris-1-Dark, MW, $N_s=14$}
\[
\Sigma=
\begin{bmatrix}
1.27 \pm 0.06 & 1.12 \pm 0.06 & 0.40 \pm 0.04\\
1.12 \pm 0.06 & 1.55 \pm 0.08 & -0.32 \pm 0.05\\
0.40 \pm 0.04 & -0.32 \pm 0.05 & 1.02 \pm 0.04\\
\end{bmatrix}
\]
\[
\mu=
\begin{bmatrix}
-0.52 \pm 0.05 & -0.64 \pm 0.06 & -0.40 \pm 0.05\\
\end{bmatrix}
\]

\subsection{Illustris-1, M31, $N_s=14$}
\[
\Sigma=
\begin{bmatrix}
1.11 \pm 0.05 & 0.88 \pm 0.05 & 0.83 \pm 0.05\\
0.88 \pm 0.05 & 1.14 \pm 0.06 & 0.08 \pm 0.05\\
0.83 \pm 0.05 & 0.08 \pm 0.05 & 1.39 \pm 0.07\\
\end{bmatrix}
\]
\[
\mu=
\begin{bmatrix}
-0.06 \pm 0.05 & -0.24 \pm 0.05 & -0.23 \pm 0.06\\
\end{bmatrix}
\]
\subsection{Illustris-1, MW, $N_s=14$}
\[
\Sigma=
\begin{bmatrix}
0.99 \pm 0.07 & 0.88 \pm 0.08 & 0.38 \pm 0.06\\
0.88 \pm 0.08 & 1.24 \pm 0.10 & -0.26 \pm 0.07\\
0.38 \pm 0.06 & -0.26 \pm 0.07 & 0.97 \pm 0.09\\
\end{bmatrix}
\]
\[
\mu=
\begin{bmatrix}
-0.39 \pm 0.05 & -0.62 \pm 0.06 & -0.28 \pm 0.05\\
\end{bmatrix}
\]

\subsection{ELVIS, M31, $N_s=14$}
\[
\Sigma=
\begin{bmatrix}
1.22 \pm 0.12 & 1.37 \pm 0.16 & 0.60 \pm 0.09\\
1.37 \pm 0.16 & 1.78 \pm 0.22 & 0.38 \pm 0.09\\
0.60 \pm 0.09 & 0.38 \pm 0.09 & 0.79 \pm 0.09\\
\end{bmatrix}
\]
\[
\mu=
\begin{bmatrix}
0.66 \pm 0.07 & 0.47 \pm 0.09 & 0.25 \pm 0.06\\
\end{bmatrix}
\]
\subsection{ELVIS, MW, $N_s=14$}
\[
\Sigma=
\begin{bmatrix}
1.83 \pm 0.19 & 1.01 \pm 0.15 & 1.42 \pm 0.16\\
1.01 \pm 0.15 & 0.92 \pm 0.13 & 0.38 \pm 0.10\\
1.42 \pm 0.16 & 0.38 \pm 0.10 & 1.64 \pm 0.20\\
\end{bmatrix}
\]
\[
\mu=
\begin{bmatrix}
-0.24 \pm 0.09 & -0.25 \pm 0.06 & -0.40 \pm 0.09\\
\end{bmatrix}
\]

\subsection{Illustris-1-Dark, M31, $N_s=15$}
\[
\Sigma=
\begin{bmatrix}
1.65 \pm 0.10 & 1.35 \pm 0.09 & 0.81 \pm 0.06\\
1.35 \pm 0.09 & 1.34 \pm 0.09 & 0.36 \pm 0.05\\
0.81 \pm 0.06 & 0.36 \pm 0.05 & 0.97 \pm 0.05\\
\end{bmatrix}
\]
\[
\mu=
\begin{bmatrix}
-0.31 \pm 0.06 & -0.51 \pm 0.05 & -0.16 \pm 0.05\\
\end{bmatrix}
\]
\subsection{Illustris-1-Dark, MW, $N_s=15$}
\[
\Sigma=
\begin{bmatrix}
1.52 \pm 0.07 & 1.25 \pm 0.06 & 0.64 \pm 0.05\\
1.25 \pm 0.06 & 1.58 \pm 0.08 & -0.15 \pm 0.06\\
0.64 \pm 0.05 & -0.15 \pm 0.06 & 1.18 \pm 0.05\\
\end{bmatrix}
\]
\[
\mu=
\begin{bmatrix}
-0.50 \pm 0.06 & -0.63 \pm 0.06 & -0.33 \pm 0.05\\
\end{bmatrix}
\]

\subsection{Illustris-1, M31, $N_s=15$}
\[
\Sigma=
\begin{bmatrix}
1.18 \pm 0.04 & 0.91 \pm 0.05 & 0.84 \pm 0.05\\
0.91 \pm 0.05 & 1.11 \pm 0.07 & 0.15 \pm 0.04\\
0.84 \pm 0.05 & 0.15 \pm 0.04 & 1.23 \pm 0.06\\
\end{bmatrix}
\]
\[
\mu=
\begin{bmatrix}
-0.25 \pm 0.06 & -0.41 \pm 0.05 & -0.32 \pm 0.06\\
\end{bmatrix}
\]
\subsection{Illustris-1, MW, $N_s=15$}
\[
\Sigma=
\begin{bmatrix}
1.06 \pm 0.07 & 0.88 \pm 0.07 & 0.44 \pm 0.06\\
0.88 \pm 0.07 & 1.11 \pm 0.08 & -0.11 \pm 0.05\\
0.44 \pm 0.06 & -0.11 \pm 0.05 & 0.84 \pm 0.08\\
\end{bmatrix}
\]
\[
\mu=
\begin{bmatrix}
-0.45 \pm 0.05 & -0.67 \pm 0.05 & -0.32 \pm 0.05\\
\end{bmatrix}
\]

\subsection{ELVIS, M31, $N_s=15$}
\[
\Sigma=
\begin{bmatrix}
1.26 \pm 0.12 & 1.25 \pm 0.13 & 0.77 \pm 0.11\\
1.25 \pm 0.13 & 1.47 \pm 0.18 & 0.43 \pm 0.10\\
0.77 \pm 0.11 & 0.43 \pm 0.10 & 1.01 \pm 0.11\\
\end{bmatrix}
\]
\[
\mu=
\begin{bmatrix}
0.48 \pm 0.08 & 0.22 \pm 0.08 & 0.23 \pm 0.07\\
\end{bmatrix}
\]
\subsection{ELVIS, MW, $N_s=15$}
\[
\Sigma=
\begin{bmatrix}
1.89 \pm 0.18 & 0.90 \pm 0.11 & 1.51 \pm 0.16\\
0.90 \pm 0.11 & 0.68 \pm 0.07 & 0.41 \pm 0.10\\
1.51 \pm 0.16 & 0.41 \pm 0.10 & 1.63 \pm 0.19\\
\end{bmatrix}
\]
\[
\mu=
\begin{bmatrix}
-0.49 \pm 0.09 & -0.55 \pm 0.06 & -0.40 \pm 0.09\\
\end{bmatrix}
\]

\section{Results from ELVIS and Illustris1}
\label{appendix:plots}

\begin{figure*}
\centering
\includegraphics[width=0.37\textwidth]{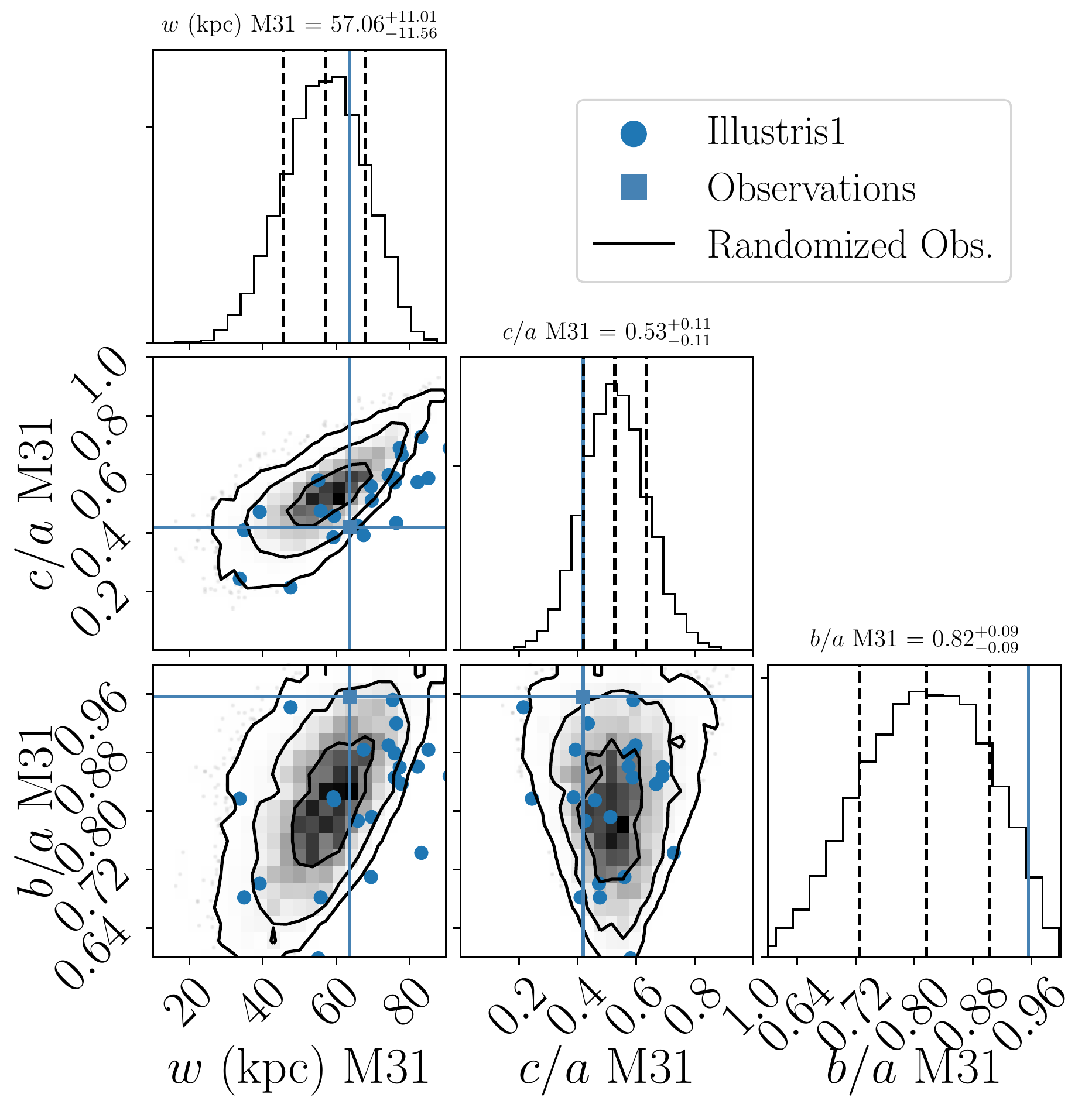}
\includegraphics[width=0.37\textwidth]{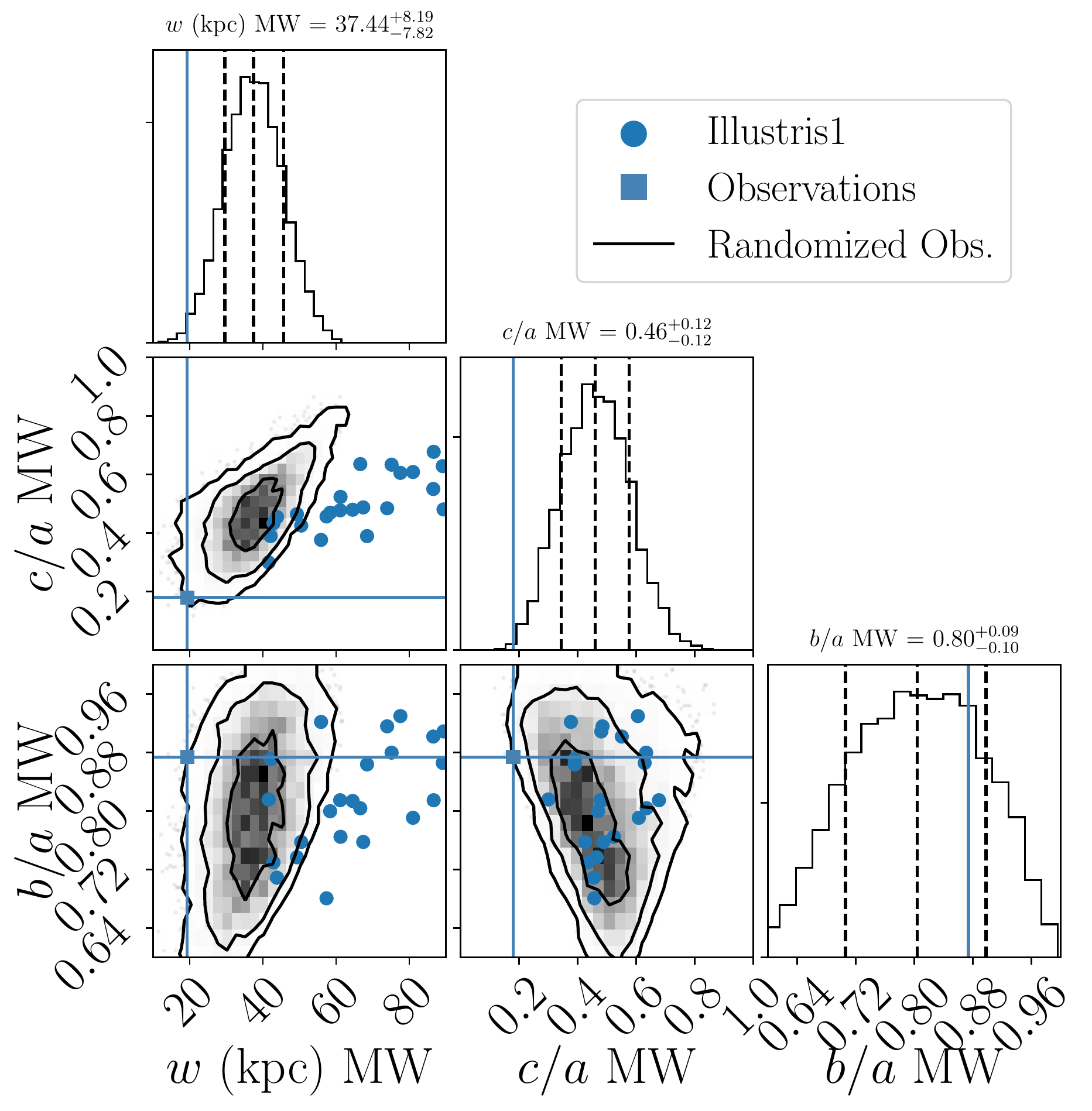}
\includegraphics[width=0.37\textwidth]{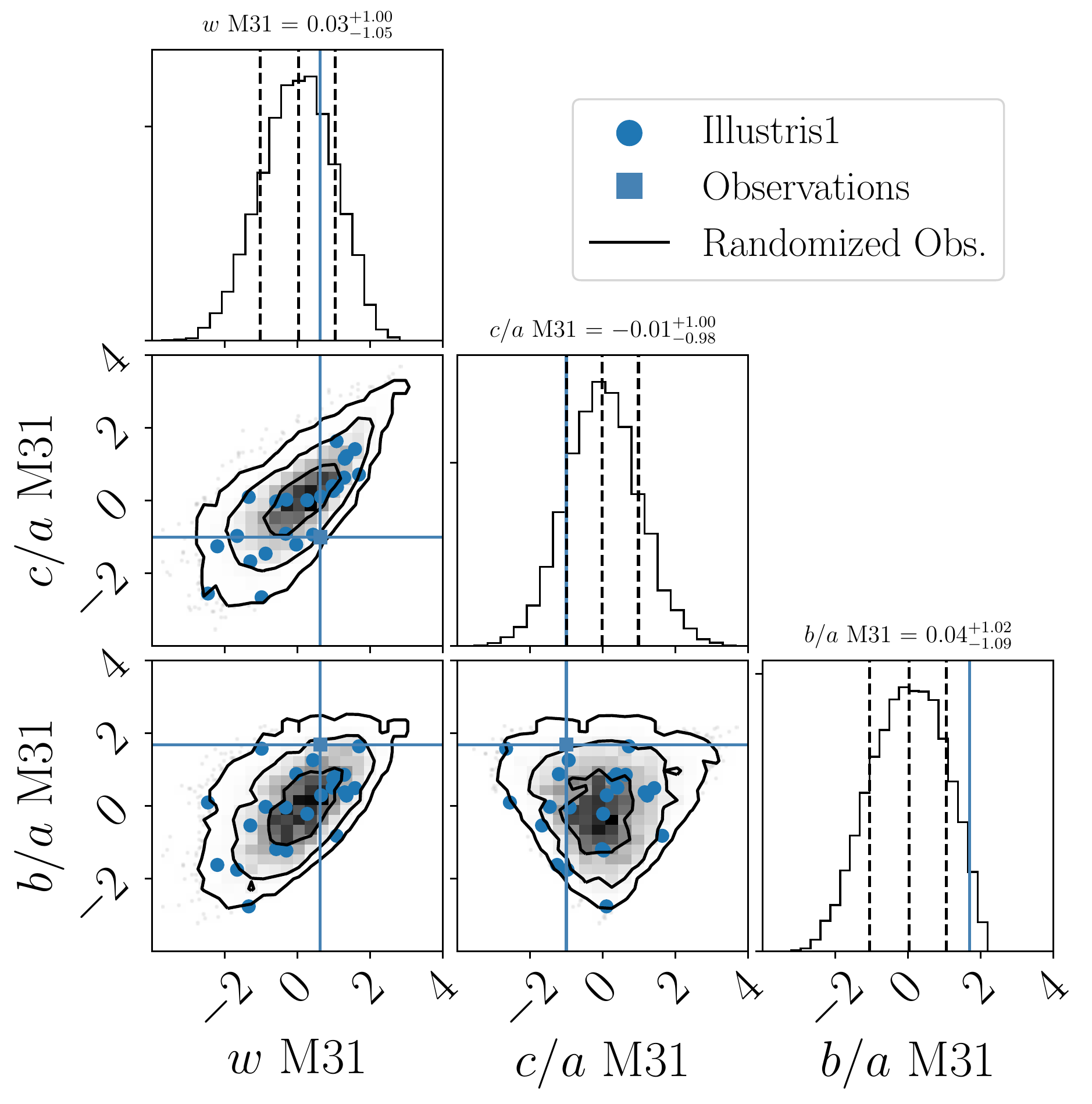}
\includegraphics[width=0.37\textwidth]{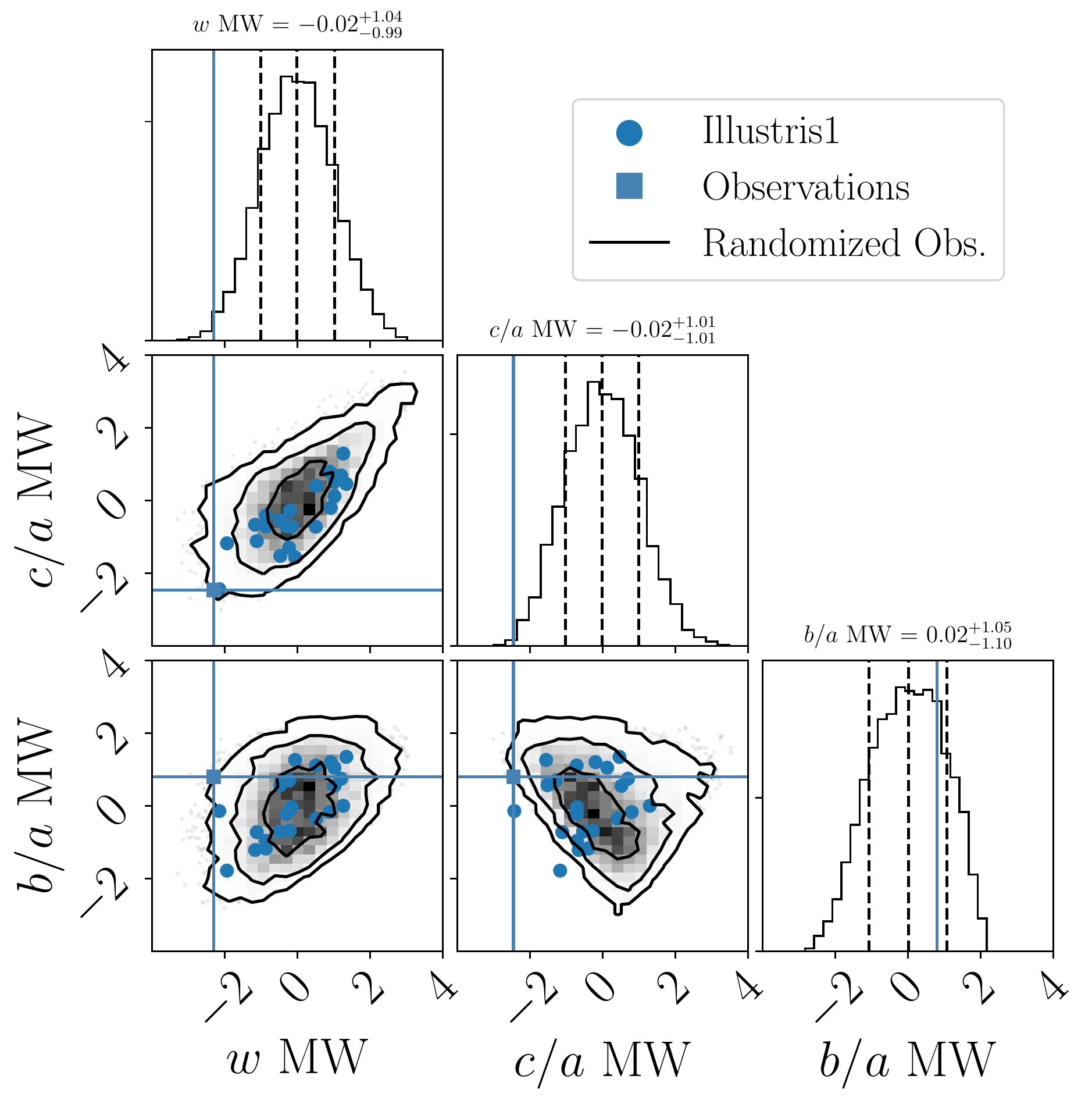}
\includegraphics[width=0.37\textwidth]{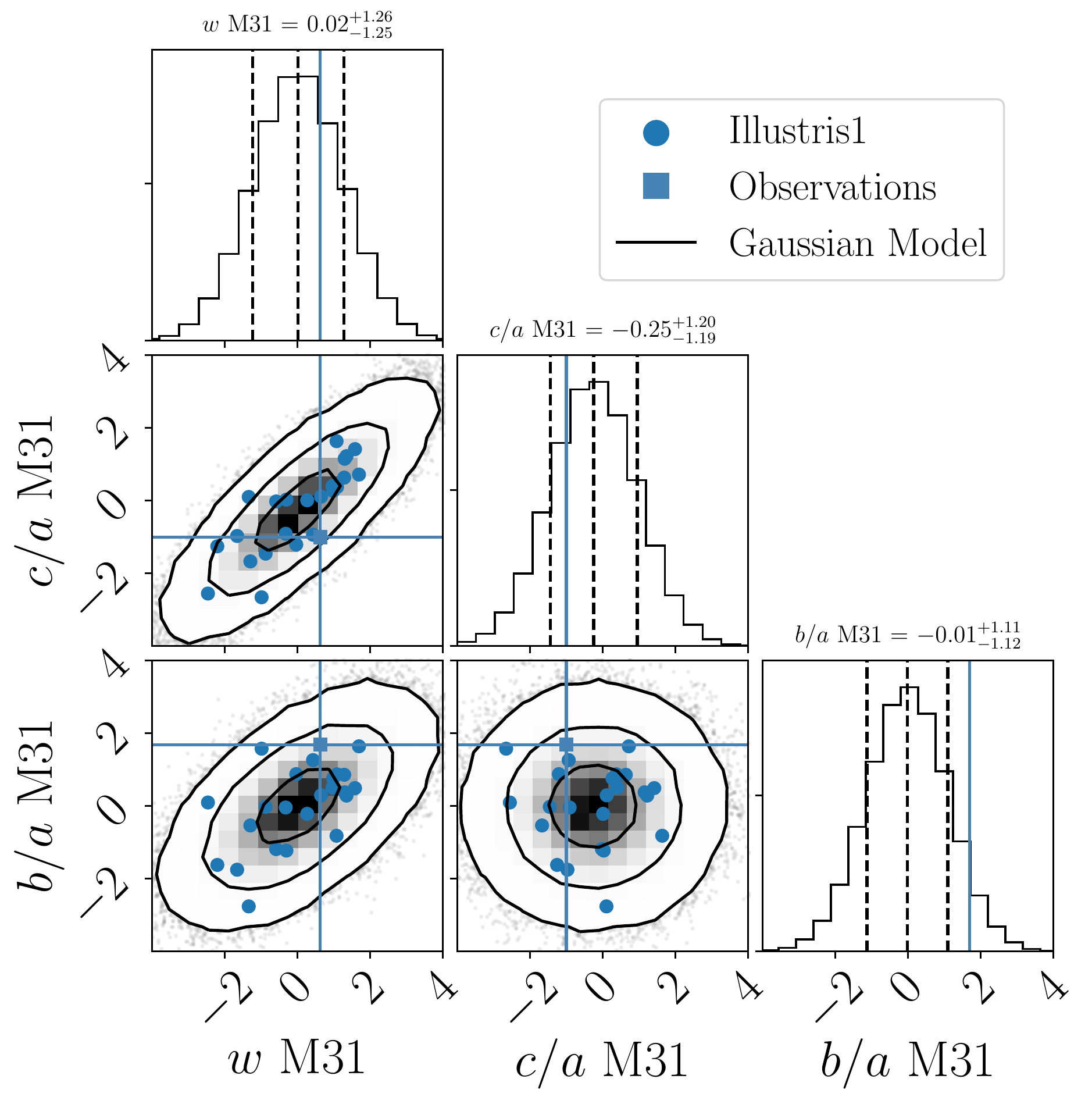}
\includegraphics[width=0.37\textwidth]{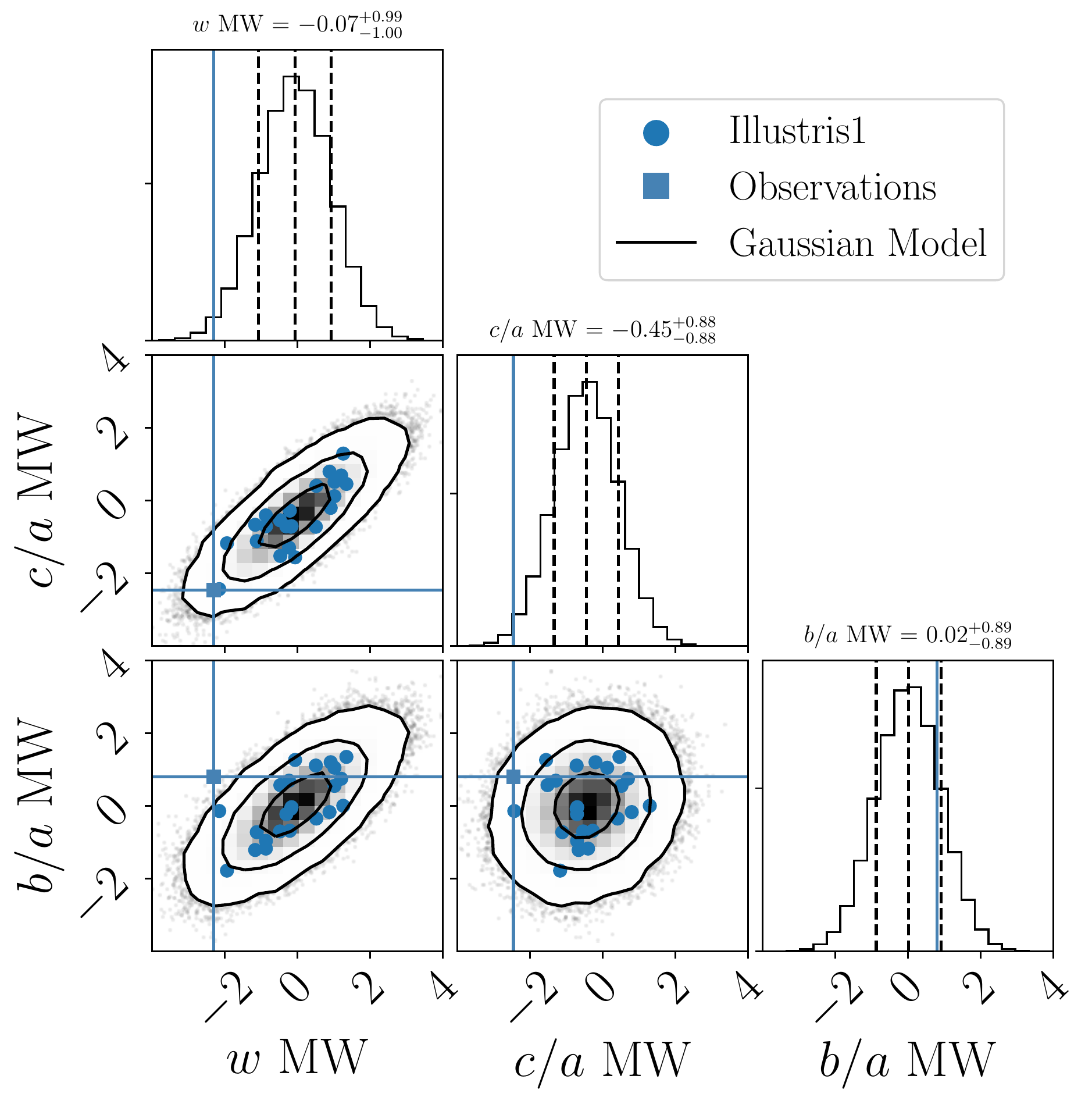}
\caption{Illustris-1 results for the same quantities presented for the Illustris-1-Dark
  simulation in Figures  
\ref{fig:physical_illustris1dark}, \ref{fig:normalized_illustris1dark}
and \ref{fig:gaussian_illustris1dark}.
Upper row corresponds to the raw values from observations and
simulated pairs, while the second row normalizes the same values to
the mean and standard deviation on its spherically randomized
counterparts. 
\label{fig:all_plots_illustris1}}
\end{figure*}

\begin{figure*}
\centering
\includegraphics[width=0.37\textwidth]{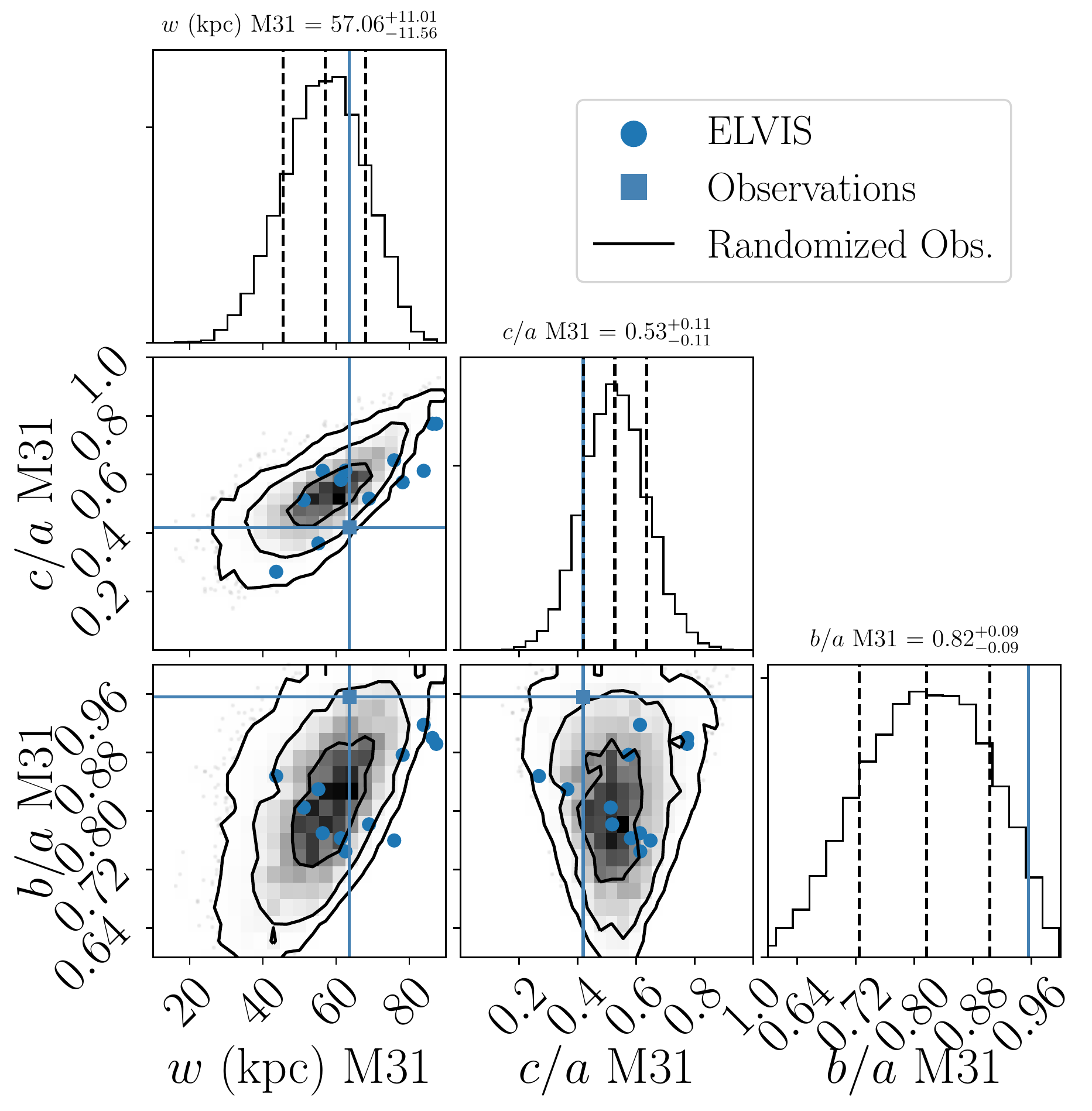}
\includegraphics[width=0.37\textwidth]{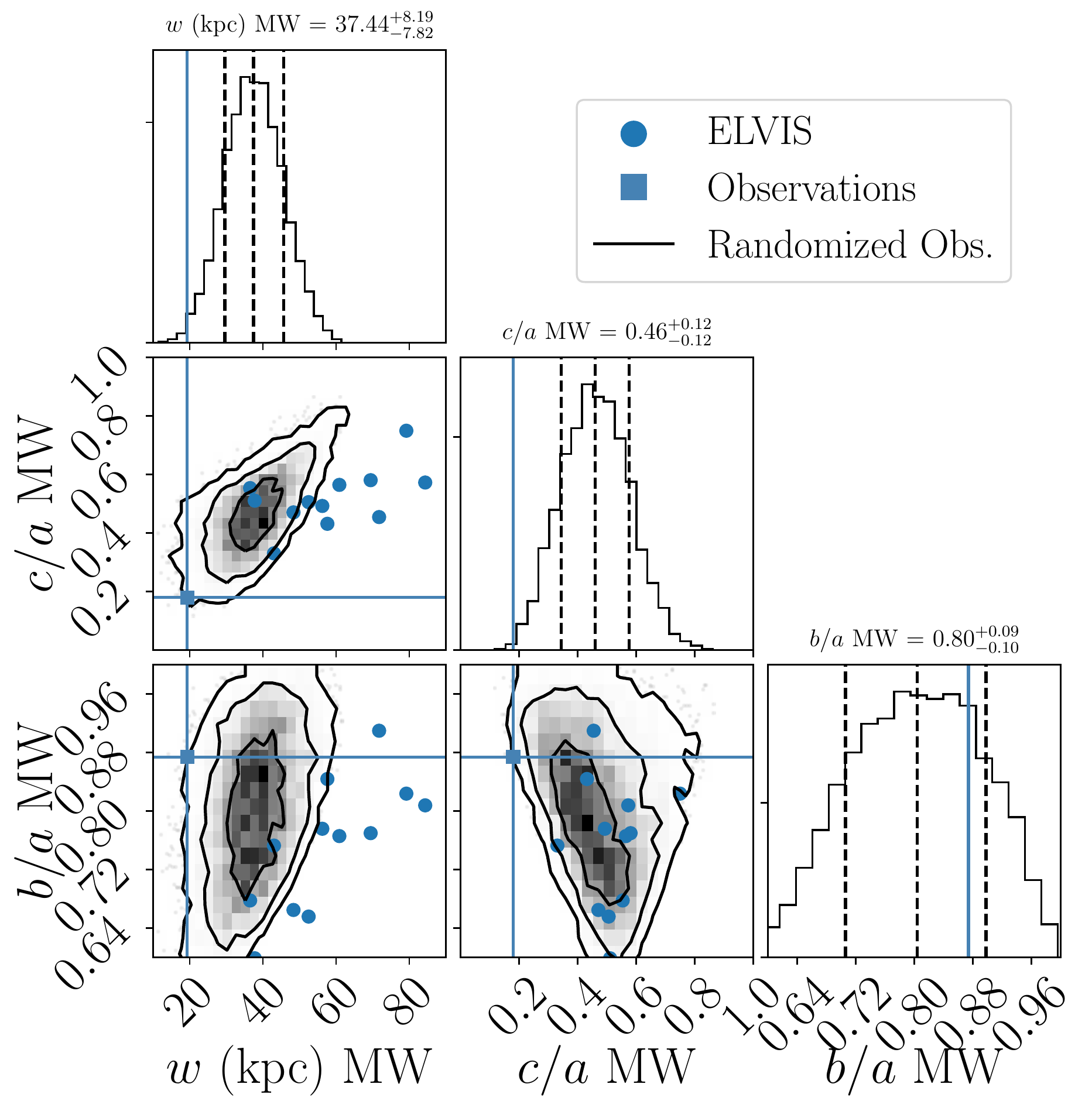}
\includegraphics[width=0.37\textwidth]{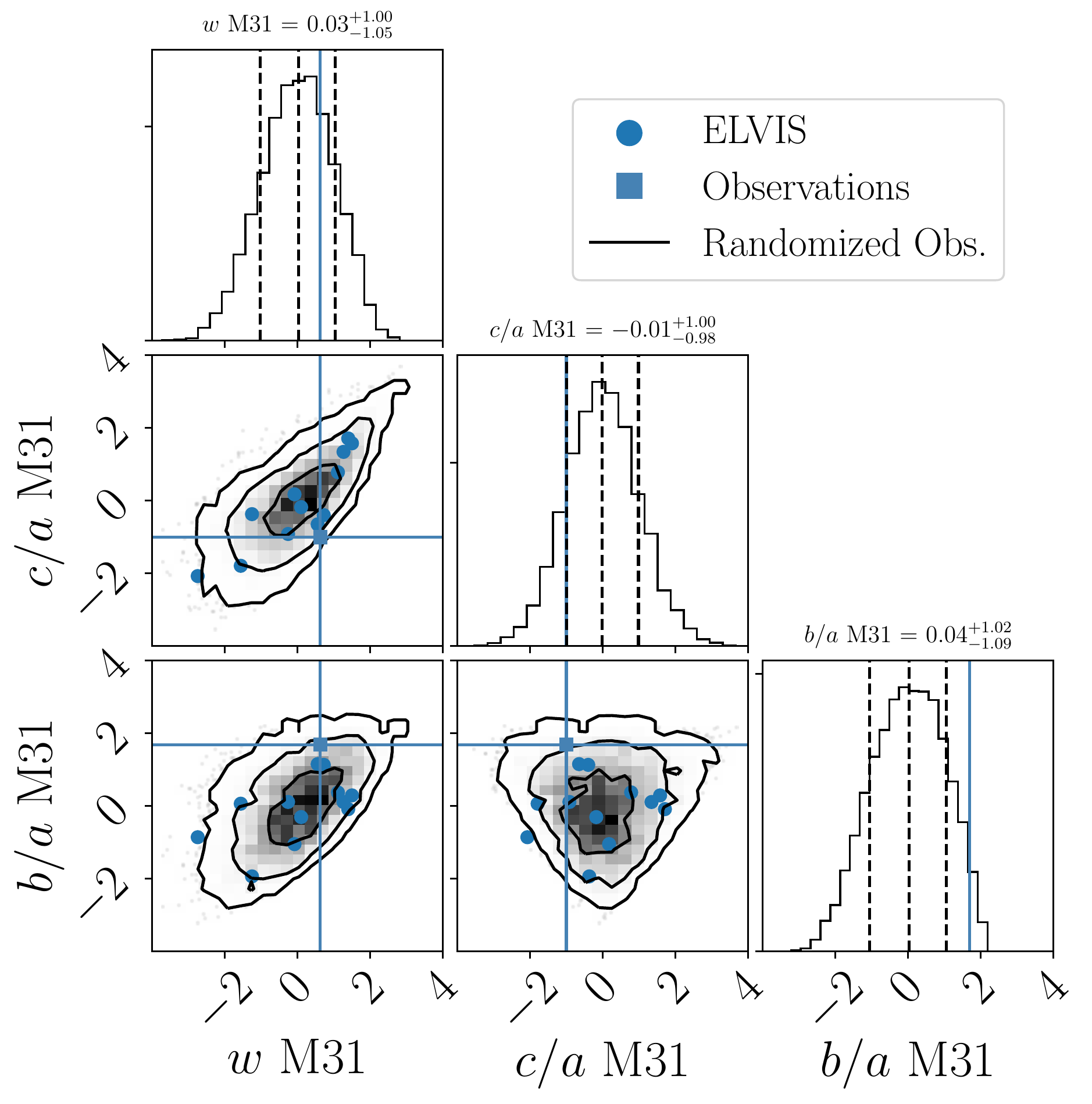}
\includegraphics[width=0.37\textwidth]{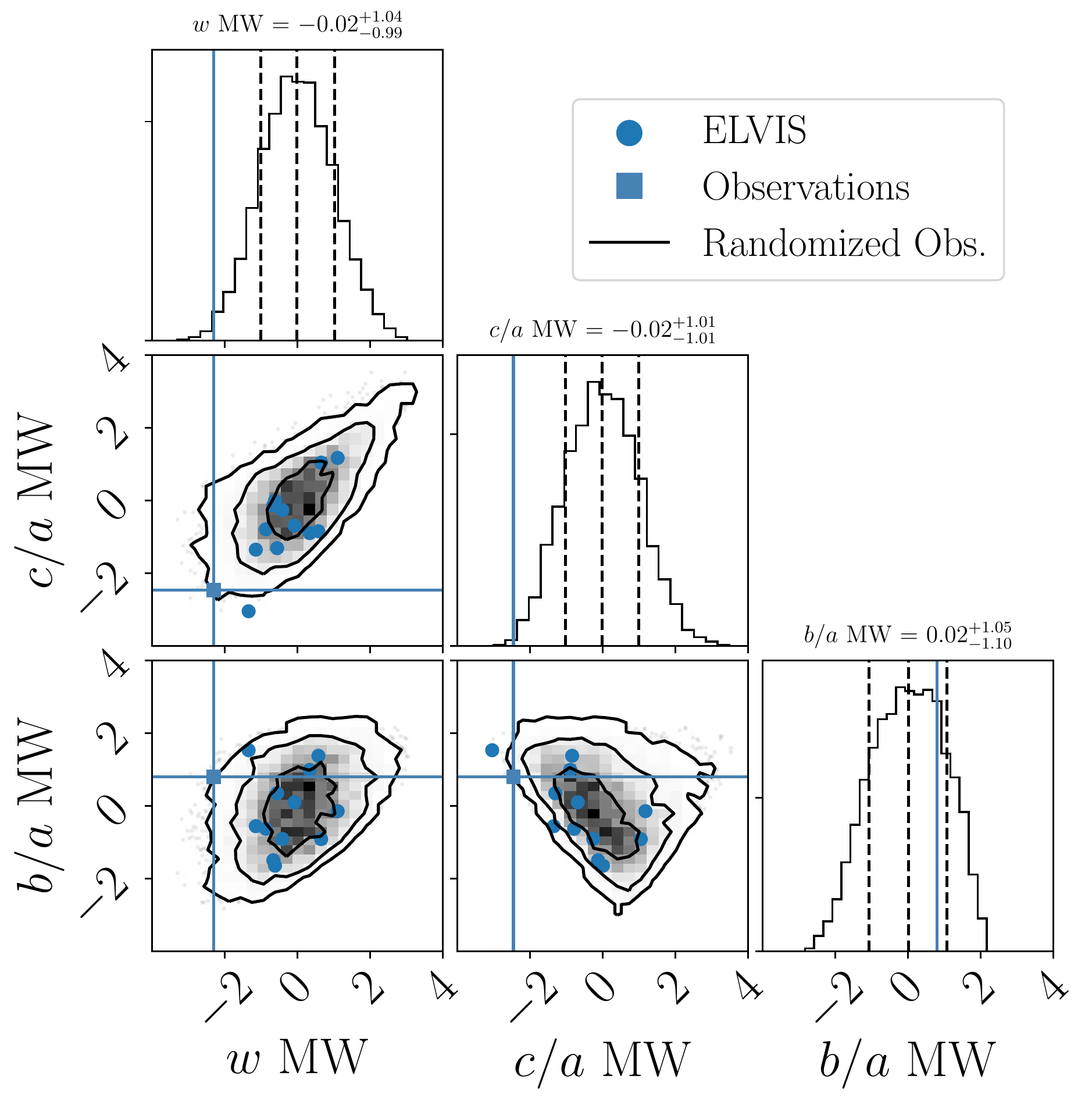}
\includegraphics[width=0.37\textwidth]{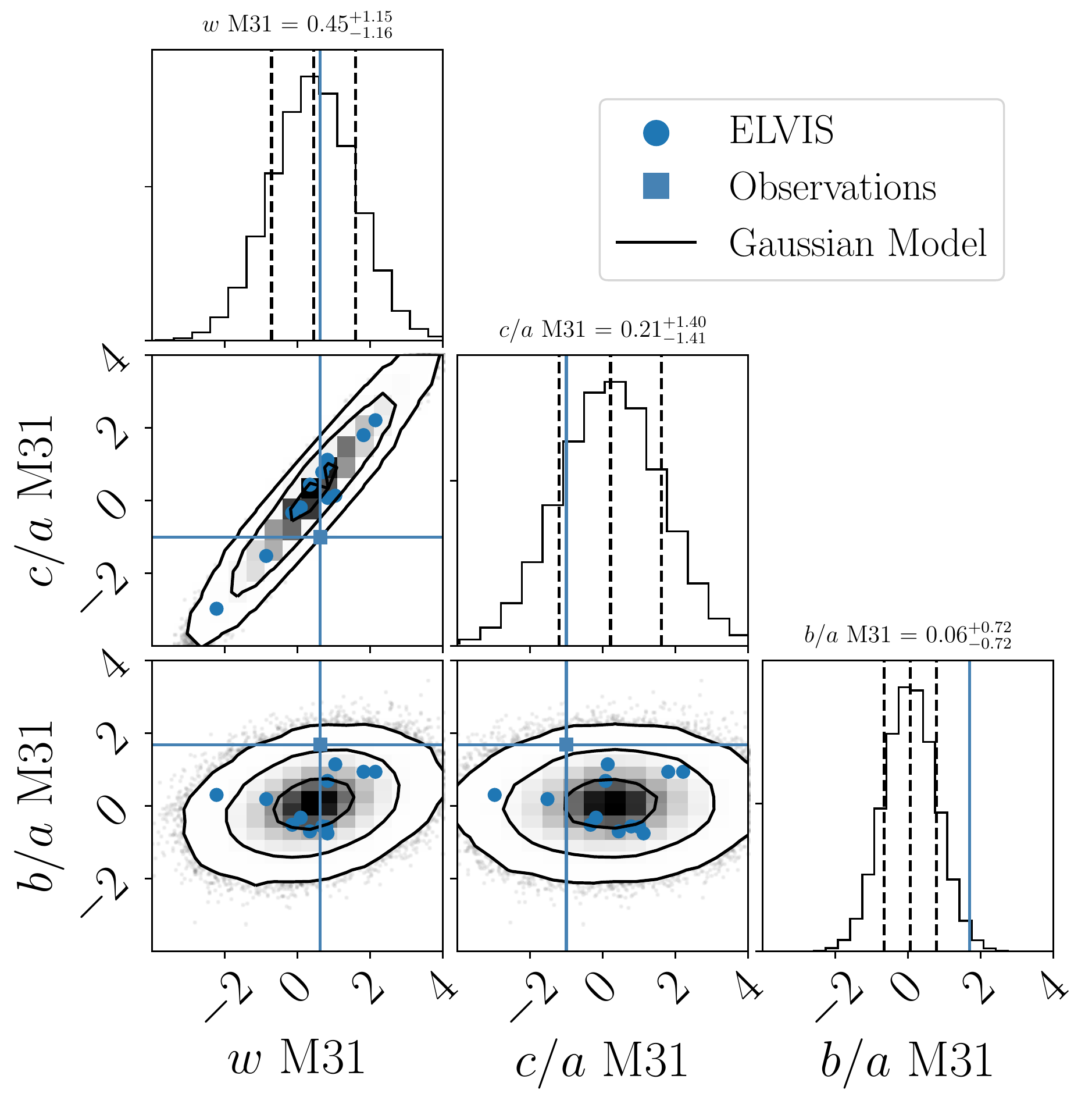}
\includegraphics[width=0.37\textwidth]{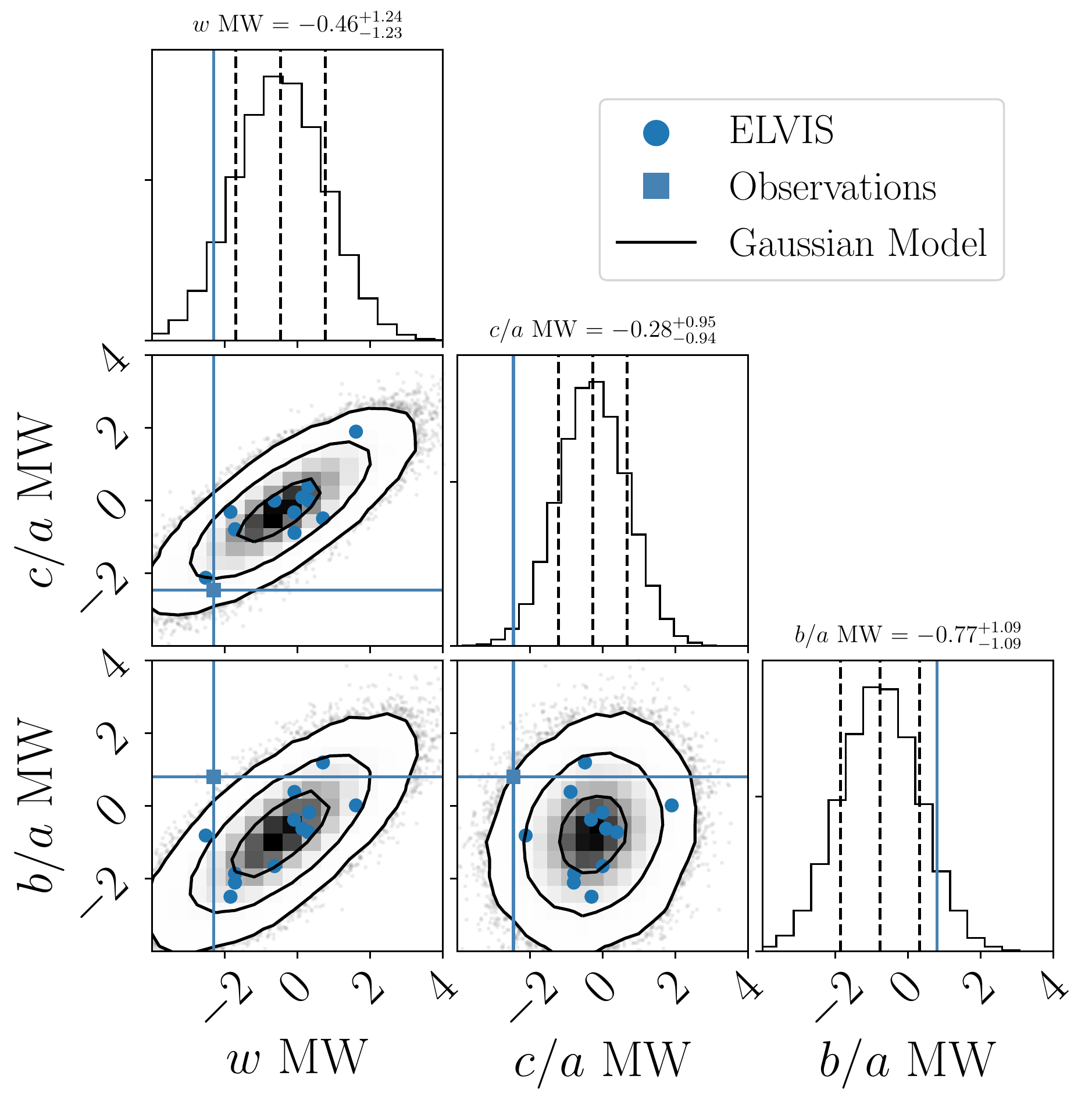}
\caption{ELVIS results for the same quantities presented for the Illustris1-Dark
  simulation in Figures  
\ref{fig:physical_illustris1dark}, \ref{fig:normalized_illustris1dark}
and \ref{fig:gaussian_illustris1dark}.
Upper row corresponds to the raw values from observations and
simulated pairs, while the second row normalizes the same values to
the mean and standard deviation on its spherically randomized
counterparts. 
\label{fig:all_plots_elvis}}
\end{figure*}

\end{document}